\RequirePackage{fix-cm}
\documentclass[smallextended]{svjour3}       
\smartqed  
\usepackage{graphicx,amssymb,amsfonts,mathrsfs,latexsym,amsmath,amssymb}
\usepackage{color}
\journalname{CCDS}
\begin{document}

\title{An explicit representation and enumeration for negacyclic codes of length $2^kn$ over $\mathbb{Z}_4+u\mathbb{Z}_4$
} \subtitle{}

\titlerunning{An explicit representation and enumeration for negacyclic codes over $\mathbb{Z}_4+u\mathbb{Z}_4$}

\author{Yonglin Cao$^1$ $\cdot$  Yuan Cao$^{1,\ 2, \ 3}$ $\cdot$ Rama Krishna Bandi$^4$ $\cdot$ Fang-Wei Fu$^{5}$
}


\institute{Yuan Cao (corresponding author) \at
             \email{yuancao@sdut.edu.cn}\\
           Yonglin Cao\at
             \email{ylcao@sdut.edu.cn}\\
           Rama Krishna Bandi\at
           bandi.ramakrishna@gmail.com\\
           Fang-Wei Fu\at
           fwfu@nankai.edu.cn\\
           $^1$School of Mathematics and Statistics, Shandong University of Technology, Zibo, Shandong 255091, China\\
           $^2$Hubei Key Laboratory of Applied Mathematics, Faculty of Mathematics and Statistics, Hubei University, Wuhan 430062, China \\
           $^3$Hunan Provincial Key Laboratory of Mathematical Modeling and Analysis in Engineering,  Changsha University of Science and Technology, Changsha, Hunan 410114, China \\
           $^4$Department of Mathematics,
International Institute of Information Technology, Naya Raipur 493661,
India\\
           $^5$Chern Institute of Mathematics and LPMC, Nankai University, Tianjin 300071, China          
          }

\date{Received: date / Accepted: date}

\maketitle

\begin{abstract}
In this paper, an explicit representation and enumeration for negacyclic codes of length $2^kn$ over the local non-principal ideal ring $R=\mathbb{Z}_4+u\mathbb{Z}_4$ $(u^2=0)$ is provided, where $k, n$ are any positive integers and $n$ is odd. As a corollary, all distinct negacyclic codes of length $2^k$ over $R$ are listed precisely. Moreover, an exact mass formula for the number of negacyclic codes of length $2^kn$ over $R$ is given and some mistakes in [Cryptogr. Commun. (2017) 9: 241--272] are corrected.

\end{abstract}

\section{Introduction}
\label{intro}
\par
    Algebraic coding theory deals with the design of error-correcting and error-detecting codes for the reliable transmission
of information across noisy channel.
   The class of constacyclic codes play a very significant role in
the theory of error-correcting codes. Since 1999, special classes of constacyclic
codes over certain classes of finite commutative chain rings have been studied
by numerous authors (see [1], [2], [5], [6], [8--10], [13], [15--17], [19-20] [22--24], for example). It is an important way and an interesting topic to construct optimal
codes (over finite fields or finite rings) from special linear codes over some appropriate rings.

\par
    Let $A$ be a finite commutative ring with identity $1\neq 0$, $A^{\times}$ the multiplicative group of units of
$A$ and $a\in
A$. We denote by $\langle a\rangle_A$ or $\langle a\rangle$ the ideal of $A$ generated by $a$, i.e. $\langle
a\rangle_A=aA$. For any ideal $I$ of $A$, we will identify the
element $a+I$ of the residue class ring $A/I$ with $a$ (mod $I$) in this paper.

   For any positive integer $N$, let
$A^N=\{(a_0,a_1,\ldots,a_{N-1})\mid a_i\in A, \ 0\leq i\leq N-1\}$ which is an $A$-module with componentwise addition and scalar multiplication by elements of $A$. Then an $A$-submodule ${\mathcal C}$ of $A^N$ is called a \textit{linear code} over $A$ of length $N$.
For any vectors $a=(a_0,a_1,\ldots,a_{N-1}), b=(b_0,b_1,\ldots,b_{N-1})\in A^N$.
The usual \textit{Euclidian inner product} of $a$ and $b$ is defined by
$[a,b]=\sum_{j=0}^{N-1}a_jb_j\in A$.
Let ${\mathcal C}$ be a linear code over $A$ of length $N$. The \textit{Euclidian dual code}
of ${\mathcal C}$ is defined by ${\mathcal C}^{\bot}=\{a\in A^N\mid [a,b]=0, \ \forall
b\in {\mathcal C}\}$, and ${\mathcal C}$ is said to be \textit{self-dual} if ${\mathcal C}={\mathcal C}^{\bot}$.

   Let $\gamma\in A^{\times}$.
Then a linear code
${\mathcal C}$ over $A$ of length $N$ is
called a $\gamma$-\textit{constacyclic code}
if $(\gamma a_{N-1},a_0,a_1,\ldots,a_{N-2})\in {\mathcal C}$ for all
$(a_0,a_1,\ldots,a_{N-1})\in{\mathcal C}$. Particularly, ${\mathcal C}$ is
a \textit{negacyclic code} if $\gamma=-1$, and ${\mathcal C}$ is
a  \textit{cyclic code} if $\gamma=1$.
  For any $a=(a_0,a_1,\ldots,a_{N-1})\in A^N$, let
$a(x)=a_0+a_1x+\ldots+a_{N-1}x^{N-1}\in A[x]/\langle x^N-\gamma\rangle$. We will identify $a$ with $a(x)$ in
this paper. It is well known that ${\mathcal C}$ is a  $\gamma$-constacyclic code
of length $N$ over $A$ if and only if ${\mathcal C}$ is an ideal of
the residue class ring $A[x]/\langle x^N-\gamma\rangle$ and that
the dual code ${\mathcal C}^{\bot}$  of a $\gamma$-constacyclic code ${\mathcal C}$ of length $N$ over
$A$ is a $\gamma^{-1}$-constacyclic code of length $N$ over
$A$, i.e. ${\mathcal C}^{\bot}$ is an ideal of $A[x]/\langle
x^N-\gamma^{-1}\rangle$ (cf. [16] Propositions 2.2 and 2.3).

\par
   In 1999, Wood in [27] showed that
for certain reasons finite Frobenius rings are the most general class of
rings that should be used for alphabets of codes. Then self-dual codes over commutative
Frobenius rings were investigated by Dougherty et al. [18]. Especially,
in 2014, codes over an extension ring of $\mathbb{Z}_4$ were studied by Yildiz et al. [28] and
[29], here the
ring  was described as $\mathbb{Z}_4[u]/\langle u^2\rangle=\mathbb{Z}_4+u\mathbb{Z}_4$ ($u^2=0$) which is a local non-principal
ring. Then a complete classification and an explicit
representation for cyclic codes of odd length over
$\mathbb{Z}_4[u]/\langle u^k\rangle=\mathbb{Z}_4+u\mathbb{Z}_4+\ldots+u^{k-1}\mathbb{Z}_4$ ($u^k=0$)
were provided by Cao et al. [7] for any integer $k\geq 2$.

\par
  Recently, Shi et al. in [25] studied $(1+2u)$-constacyclic codes over the ring
$\Gamma=\mathbb{Z}_{4}[u]/\langle u^2-1\rangle=\mathbb{Z}_{4}+u\mathbb{Z}_{4}$ ($u^2=1$) of odd length $n$, and
investigated properties of these codes and their $\mathbb{Z}_{4}$ images. As a further development, a complete classification for simple-root cyclic codes over non-principal rings $\mathbb{Z}_{p^s}[v]/\langle v^2-pv\rangle$ was presented in [11] for any prime number $p$ and integer $s\geq 2$. Recently,
negacyclic codes over the local ring $\mathbb{Z}_4[v]/\langle v^2+2v\rangle$
of oddly even length and their Gray images were studied in [12].

\par
  In 2017, Bandi et al. [4] studied negacyclic codes of length $2^k$ over
$R=\mathbb{Z}_{4}+u\mathbb{Z}_{4}$ ($u^2=0$). Some results in [4] are the following:

\par
  ($\dag$) \textsl{Described the ideal structure of $\frac{R[x]}{\langle x^{2^k}+1\rangle}$ (distinct negacyclic codes
 of length $2^k$  over $R$) by Theorem 11 in Pages 248--249 of [4], and given a mass formula for the number of negacylic codes of length $2^k$ over $R$ by Theorem 12 in Pages 249 of [4]. The proof for this mass formula takes $11$ pages: Pages 250--260.}

\par
  ($\ddag$) \textsl{Described roughly the ideal structure of $\frac{R[x]}{\langle x^{2^ke}+1\rangle}$, where $e$ is an odd integer,
by Theorems 18--20, discrete Fourier transform and Theorem 21 in Pages 268--270 of [4]. Moreover, by Theorem 22 of Page 270 in [4] the number of distinct negacyclic codes of length $2^ke$ over $R$ is given by}
$\prod_{\xi\in J}N_\xi$,
\textsl{where}

\par
  $\triangleright$ \textsl{$J$ denotes a complete set of representatives of
the $2$-cyclotomic cosets modulo $e$ (see Page 270 in [4])}.

\par
  $\triangleright$ \textsl{For each $\xi\in J$, $r_\xi$ is the size of the
$2$-cyclotomic cosets modulo $e$ containing $\xi$, ${\rm GR}(R,r_\xi)$ is a Galois ring extension of
$R$ with degree $r_\xi$ (see Page 267 in [4]) and $N_\xi$ is the number of distinct ideals in the ring
$S_{r_\xi}=\frac{{\rm GR}(R,r_\xi)[x]}{\langle x^{2^k}+1\rangle}$ (see Pages 267 and 270 in [4])}.

\par
   There are three problems in ref. [4]:

\par
   $\diamondsuit$ The mass formula for the number of negacylic codes of length $2^k$ over $R$ given by Theorem 12 in [4]
is wrong (see Remark 4.4 in this paper);

\par
   $\diamondsuit$ The number $N_\xi$ of distinct ideals in the ring
$S_{r_\xi}=\frac{{\rm GR}(R,r_\xi)[x]}{\langle x^{2^k}+1\rangle}$ had not been determined for any $\xi\in J$ in [4].

\par
  Hence an explicit formula for the number of all distinct negacyclic codes over $\mathbb{Z}_{4}+u\mathbb{Z}_{4}$ of arbitrary even length has not been obtained.

\par
  $\diamondsuit$ Although negacyclic codes over $\mathbb{Z}_{4}+u\mathbb{Z}_{4}$ of arbitrary even length
were studied in Section 4 of [4], but the expression for each code, given
by Theorems 18--20, discrete Fourier transform and Theorem 21 in Pages 268--270 of [4],
is a little complicated and not clear enough to list all distinct negacyclic codes over $\mathbb{Z}_{4}+u\mathbb{Z}_{4}$
of specific even length.

\par
  Motivated by those, we adopt a new idea and use some new methods to study
negacyclic codes over $\mathbb{Z}_{4}+u\mathbb{Z}_{4}$ of arbitrary even length.

\par
   In this paper, let
$\mathbb{Z}_{4}=\{0,1,2,3\}$ in which the arithmetic is done
modulo $4$
and denote $\mathbb{F}_2=\{0,1\}$ in which the arithmetic is done modulo $2$. We will regard $\mathbb{F}_2$ as a subset of $\mathbb{Z}_{4}$ in this paper. But $\mathbb{F}_2$ is not a subring of $\mathbb{Z}_{4}$.
 Let $a\in\mathbb{Z}_{4}$. Then $a$ has a unique $2$-adic expansion: $a_0+2a_1$, $a_0,a_1\in \mathbb{F}_2$.
It is well known that $a\in \mathbb{Z}_{4}^{\times}$
if and only if $a_0\neq 0$.
Denote $\overline{a}=a_0\in \mathbb{F}_2$. Then
$^{-}: a\mapsto \overline{a}$ ($\forall a\in \mathbb{Z}_{4}$) is a ring homomorphism from
$\mathbb{Z}_{4}$ onto $\mathbb{F}_2$, and this homomorphism can be extended to a ring homomorphism from
$\mathbb{Z}_{4}[y]$ onto $\mathbb{F}_2[y]$ by:
$$\overline{f}(y)=\sum_{i=0}^m\overline{b}_iy^i, \
\forall f(y)=\sum_{i=0}^mb_iy^i\in \mathbb{Z}_{4}[y].$$
A monic polynomial $f(y)\in \mathbb{Z}_{4}[y]$
of positive degree is said to be \textit{basic irreducible} if $\overline{f}(y)$ is an irreducible
polynomial in $\mathbb{F}_2[y]$ (cf. $\S$ 13.4 in [26]).

\par
  For any monic polynomial
$f(y)\in \mathbb{Z}_{4}[y]$ of degree $d\geq 1$, let $\frac{\mathbb{Z}_{4}[y]}{\langle f(y)\rangle}=\{\sum_{i=0}^{d-1}a_iy^i\mid
a_0,a_1,\ldots,a_{d-1}\in \mathbb{Z}_4\}$ in which the arithmetic is done modulo $f(y)$ and
$\frac{\mathbb{F}_{2}[y]}{\langle \overline{f}(y)\rangle}=\{\sum_{i=0}^{d-1}b_iy^i\mid
b_0,b_1,\ldots,b_{d-1}\in \mathbb{F}_2\}$ in which the arithmetic is done modulo $\overline{f}(y)$.
In this paper, we still use $^{-}$ to denote the homomorphism of
rings from $\frac{\mathbb{Z}_{4}[y]}{\langle f(y)\rangle}$ onto $\frac{\mathbb{F}_{2}[y]}{\langle \overline{f}(y)\rangle}$ defined by:   $$\sum_{i=0}^{d-1}a_iy^i\mapsto \sum_{i=0}^{d-1}\overline{a}_iy^i, \ \forall a_0, a_1, \ldots, a_{d-1}\in \mathbb{Z}_{4}.$$

\par
   In this paper, let $k$ be any positive integer and $n$ be an odd positive integer. We assume
\begin{equation}
y^n-1=f_1(y)f_2(y)\ldots f_r(y),
\end{equation}
 where $f_1(y), f_2(y),\ldots, f_r(y)$ are pairwise coprime monic basic irreducible polynomials in $\mathbb{Z}_{4}[y]$ and
 $${\rm deg}(f_j(y))=d_j, \ j=1,\ldots,r.$$
 Then $\overline{f}_1(y)$, $\overline{f}_2(y)$, $\ldots$, $\overline{f}_r(y)$ are pairwise coprime irreducible polynomials in $\mathbb{F}_{2}[y]$
and ${\rm deg}(\overline{f}_j(y))=d_j$ for all $j=1,\ldots,r$.
We will
adopt the following notation.

\vskip 2mm\par
  $\bullet$ $\mathbb{Z}_{4}+u\mathbb{Z}_{4}=\mathbb{Z}_{4}[u]/\langle u^2\rangle=\{a+ub\mid a,b\in \mathbb{Z}_{4}\}$
$(u^2=0)$ in which the operations are defined by
$$\alpha+\beta=(a+b)+u(c+d) \ {\rm and} \ \alpha\beta=ac+u(ad+bc),$$
for any $\alpha=a+bu,\beta=c+du\in \mathbb{Z}_{4}+u\mathbb{Z}_{4}$ with $a,b,c,d\in \mathbb{Z}_{4}$. Then
$\mathbb{Z}_{4}+u\mathbb{Z}_{4}$ is a local non-principal ideal ring (cf. [29]).

\par
  $\bullet$ $\mathcal{A}=\frac{\mathbb{Z}_{4}[x]}{\langle x^{2^kn}+1\rangle}=\{\sum_{i=0}^{2^kn-1}a_ix^i\mid
a_0, a_1,\ldots, a_{2^kn-1}\in \mathbb{Z}_{4}\}$ in which the arithmetic is done modulo $x^{2^kn}+1$.

\par
  Let $1\leq j\leq r$ in the following.

\par
  $\bullet$ $\mathcal{R}_j=\frac{\mathbb{Z}_{4}[x]}{\langle f_j(-x^{2^k})\rangle}=
\{\sum_{i=0}^{2^kd_j-1}a_ix^i\mid a_0,a_1,\ldots,a_{2^kd_j-1}\in
\mathbb{Z}_{4}\}$ in which the arithmetic is done modulo $f_j(-x^{2^k})$, where
${\rm deg}(f_j(-x^{2^k}))=2^kd_j$.

\par
  $\bullet$ $\mathcal{T}_j=
\{\sum_{i=0}^{d_j-1}t_ix^i\mid t_0,t_1,\ldots,t_{d_j-1}\in
\mathbb{F}_2\}\subset \mathcal{R}_j$ with $|\mathcal{T}_j|=2^{d_j}$.

\vskip 3mm\par
    The present paper is organized as follows. In Section 2, we prove that each
$\mathcal{R}_j$ is a finite chain ring, $1\leq j\leq r$, and establish an explicit isomorphism
of rings from the direct product ring $\mathcal{R}_1\times \ldots\times \mathcal{R}_r$
onto $\mathcal{A}$.
In Section 3, we construct a precise isomorphism of rings from the direct
product ring $(\mathcal{R}_1+u\mathcal{R}_1)\times\ldots\times (\mathcal{R}_r+u\mathcal{R}_r)$ onto
$\frac{(\mathbb{Z}_{4}+u\mathbb{Z}_{4})[x]}{\langle x^{2^kn}+1\rangle}$ first. Then we present all distinct
ideals of each ring $\mathcal{R}_j+u\mathcal{R}_j$ explicitly.
Hence we give an explicit representation and enumeration for negacyclic codes of length $2^kn$ over $\mathbb{Z}_4+u\mathbb{Z}_4$.
In Section 4, we list all distinct negacyclic codes of length $2^k$ over $\mathbb{Z}_4+u\mathbb{Z}_4$ by their explicit expressions, obtain an exact formula to count the number of these codes and correct
a mistake for the mass formula to the number of negacyclic codes of length $2^k$ over $\mathbb{Z}_4+u\mathbb{Z}_4$ obtained in [4, Cryptogr. Commun. (2017) 9: 241--272]. In Section 5, we give an explicit representation for
all distinct cyclic codes of odd length $n$ over $\mathbb{Z}_4+u\mathbb{Z}_4$. Section 6 concludes the paper.

\section{Structure of the ring $\mathcal{A}=\frac{\mathbb{Z}_{4}[x]}{\langle x^{2^kn}+1\rangle}$}
In this section, we consider how to decompose the ring $\mathcal{A}$
into a direct sum of finite chain rings first.

\vskip 3mm \noindent
  {\bf Lemma 2.1} ([15] Proposition 2.1) \textit{Let $A$ be a finite associative and commutative
  ring with identity. Then the following conditions are equivalent}:

\par
   (i) \textit{$A$ is a local ring and the maximal ideal $M$ of $A$ is principal, i.e., $M=\langle \pi\rangle$
for some $\pi\in A$};

\par
   (ii) \textit{$A$ is a local principal ideal ring};

\par
   (iii) \textit{$A$ is a chain ring with ideals $\langle \pi^i\rangle$, $0\leq i\leq \nu$,
where $\nu$ is the nilpotency index of $\pi$}.

\vskip 3mm \noindent
  {\bf Lemma 2.2} ([20] Proposition 2.2) \textit{Let $A$ be a finite commutative chain ring,
with maximal ideal $M=\langle\pi\rangle$, and let $\nu$ be the
nilpotency index of $\pi$. Then}
\par
   (i) \textit{For some prime $p$ and positive integer $m$, $|A/\langle \pi\rangle|=q$ where $q=p^m$,
$|A|=q^{\nu}$, and the characteristic of $A/\langle \pi\rangle$ and
$A$ are powers of $p$};
\par
  (ii) \textit{For $i=0,1,\ldots,\nu$, $|\langle\pi^i\rangle|=q^{\nu-i}$}.

\vskip 3mm\noindent
    {\bf Lemma 2.3} ([20] Lemma 2.4)\textit{Using the notations in Lemma 2.2, let
$V\subseteq A$ be a system of representatives for the equivalence classes of $A$ under
congruence modulo $\pi$. (Equivalently, we can define $V$ to be a maximal subset of $A$ with the property that
$r_1-r_2\not\in \langle \pi\rangle$ for all $r_1,r_2\in V$, $r_1\neq r_2$.) Then}

\vskip 2mm \par
  (i) \textit{Every element $a$ of $A$ has a unique $\pi$-adic expansion: $a=\sum_{j=0}^{\nu-1}r_j\pi^j$,
$r_0,r_1,\ldots$, $r_{\nu-1}\in V$}.

\vskip 2mm \par
  (ii) \textit{$|A/\langle \pi\rangle|=|V|$ and $|\langle \pi^i\rangle|=|V|^{\nu-i}$ for
$0\leq i\leq \nu-1$}.

\vskip 3mm \par
  Let $1\leq j\leq r$. From now on, we adopt the following notation.

\par
  $\bullet$ $\Gamma_j=\frac{\mathbb{Z}_{4}[y]}{\langle f_j(y)\rangle}=\{\sum_{i=0}^{d_j-1}a_iy^i\mid a_0,a_1,\ldots,a_{d_j-1}\in
\mathbb{Z}_{4}\}$ in which the arithmetic is done modulo $f_j(y)$.

\par
  $\bullet$ $\overline{\Gamma}_j=\frac{\mathbb{F}_{2}[y]}{\langle \overline{f}_j(y)\rangle}=\{\sum_{i=0}^{d_j-1}b_iy^i\mid b_0,b_1,\ldots,b_{d_j-1}\in
\mathbb{F}_{2}\}$ in which the arithmetic is done modulo $\overline{f}_j(y)$.

\vskip 3mm \noindent
   {\bf Lemma 2.4}
  (i)  (cf. [26] Theorem 14.1]) \textit{$\Gamma_j$ is a Galois ring of characteristic $4$ and cardinality $4^{d_j}$,
in symbol as $\Gamma_j={\rm GR}(4,4^{d_j})$. Moreover, we have
$\Gamma_j=\mathbb{Z}_{4}[\zeta_j]$, where $\zeta_j=y\in \Gamma_j$ satisfying
$\zeta_j^{2^{d_j}-1}=1$, i.e. $\zeta_j^{2^{d_j}}=\zeta_j$}.

\par
   \textit{Denote $\overline{\zeta}_j=y\in \overline{\Gamma}_j$. Then $\overline{\Gamma}_j=\mathbb{F}_2[\overline{\zeta}_j]$ which is a finite field of cardinality $2^{d_j}$,
$\overline{f}_j(x)=\prod_{i=0}^{d_j-1}(x-\overline{\zeta}_j^{2^i})\in \mathbb{F}_2[x]$ and that $^{-}$ can be extended to a ring homomorphism from
$\Gamma_j$ onto $\overline{\Gamma}_j$ by $\xi\mapsto \overline{\xi}=\sum_{i=0}^{d_j-1}\overline{a}_i\overline{\zeta}_j^i$,
for all $\xi=\sum_{i=0}^{d_j-1}a_i\zeta_j^i\in \Gamma_j$ where $a_0,a_1,\ldots,a_{d_j-1}\in \mathbb{Z}_{4}$}.

\par
  (ii) (cf. [6] Lemma 2.3(ii)) \textit{$f_j(x)=\prod_{i=0}^{d_j-1}(x-\zeta_j^{2^i})$ in $\Gamma_j[x]$}.

\vskip 3mm\par
  Now, we determine the algebraic structure of each ring $\mathcal{R}_j$, $1\leq j\leq r$.

\vskip 3mm\noindent
   {\bf Lemma 2.5}  \textit{Using the notation in Section 1, let $1\leq j\leq r$. Then}

\vskip 2mm\par
   (i) \textit{There is an invertible element $\vartheta_j(x)$ of the ring $\mathcal{R}_j$ such that
$f_j(x)^{2^k}=2\vartheta_j(x)$ in $\mathcal{R}_j$. Hence
$\langle 2\rangle=\langle f_j(x)^{2^k}\rangle$ as ideals of $\mathcal{R}_j$}.

\vskip 2mm\par
   (ii) \textit{${\mathcal R}_j$ is a finite chain ring with the unique maximal ideal $\langle f_j(x)\rangle$
generated by $f_j(x)$ where $\langle f_j(x)\rangle=f_j(x){\mathcal R}_j$, the nilpotency
index of $f_j(x)$ is equal to $2^{k+1}$ and ${\mathcal R}_j/\langle f_j(x)\rangle$ is a finite field of cardinality $2^{d_j}$}.

\vskip 2mm\par
   (iii) \textit{Each element $\alpha\in {\mathcal R}_j$ has a unique $f_j(x)$-adic expansion}:
$$\alpha=b_0(x)+b_1(x)f_j(x)+\ldots+b_{2^{k+1}-1}(x)f_j(x)^{2^{k+1}-1},$$
\textit{where $b_i(x)\in {\mathcal T}_j$ for all $i=0,1,\ldots,2^{k+1}-1$}.

\vskip 2mm \par
  (iv) \textit{All distinct ideals of $\mathcal{R}_j$ are given by: $\langle f_j(x)^l\rangle= f_j(x)^l\mathcal{R}_j$,
$l=0,1,2,$ $\ldots,2^{k+1}$. Moreover, we have $|\langle f_j(x)^l\rangle|=2^{(2^{k+1}-l)d_j}$}.

\vskip 2mm \par
  (v) \textit{Let $1\leq l\leq 2^{k+1}$. Then $\frac{\mathcal{R}_j}{\langle f_j(x)^l\rangle}=\{\sum_{i=0}^{l-1}b_i(x)f_j(x)^i\mid
b_0(x),\ldots,b_{l-1}\in \mathcal{T}_j\}$ $($in which $f_j(x)^l=0)$ and $|\frac{\mathcal{R}_j}{\langle f_j(x)^l\rangle}|=2^{ld_j}$}.

\vskip 2mm\par
  (vi) \textit{Let $0\leq l\leq t \leq 2^{k+1}-1$. Then}
$f_j(x)^l\cdot\frac{\mathcal{R}_j}{\langle f_j(x)^t\rangle}=\{\sum_{i=l}^{t-1}b_i(x)f_j(x)^i\mid
b_l(x),\ldots,b_{t-1}\in \mathcal{T}_j\},$
\textit{where we set
$f_j(x)^l\cdot\frac{\mathcal{R}_j}{\langle f_j(x)^l\rangle}=\{0\}$ for
convenience, and we have $|f_j(x)^l\cdot \frac{\mathcal{R}_j}{\langle f_j(x)^t\rangle}|=2^{(t-l)d_j}$}.

\vskip 3mm\noindent
   \textit{Proof} (i) By Lemma 2.4(ii), we have $f_j(x)=\prod_{i=0}^{d_j-1}(x-\zeta_j^{2^i})$ in $\Gamma_j[x]$.
This implies $f(x)^{2^k}=\prod_{i=0}^{d_j-1}(x-\zeta_j^{2^i})^{2^k}$, where
$$\left(x-\zeta_j^{2^i}\right)^{2^k}
=\left(-(-x^{2^k}-(\zeta_j^{2^i})^{2^k})+\left(\begin{array}{c}2^k\cr 2^{k-1}\end{array}\right)x^{2^{k-1}}(-\zeta_j^{2^i})^{2^{k-1}}\right)$$
since $\left(\begin{array}{c}2^k\cr 2^{k-1}\end{array}\right)=2$ and $\left(\begin{array}{c}2^k\cr t\end{array}\right)=0$ in $\mathbb{Z}_4$ for all $t\not\in \{0,2^{k-1},2^k\}$. Hence
\begin{eqnarray*}
f(x)^{2^k}
 &=& (-1)^{d_j}\prod_{i=0}^{d_j-1}\left((-x^{2^k})-\zeta_j^{2^{i+k}}\right)\\
  && +2\prod_{i=0}^{d_j-1}x^{2^{k-1}}(-\zeta_j^{2^i})^{2^{k-1}}
  \prod_{0\leq t\leq d_j-1, t\neq i}\left(x^{2^k}+(\zeta_j^{2^t})^{2^k}\right).
\end{eqnarray*}
Denote $g_j(x)=\prod_{i=0}^{d_j-1}x^{2^{k-1}}(-\zeta_j^{2^i})^{2^{k-1}}
  \prod_{0\leq t\leq d_j-1, t\neq i}(x^{2^k}+(\zeta_j^{2^t})^{2^k})\in \Gamma_j[x]$. Then we have
  $$f(x)^{2^k}=(-1)^{d_j}f_j(-x^{2^k})+2g_j(x).$$
This implies
$g_j(x)=\frac{f(x)^{2^k}-(-1)^{d_j}f_j(-x^{2^k})}{2}$ (mod $4$) and so $g_j(x)\in \mathbb{Z}_4[x]$.
As $4=0$, we have $2g(x)=2\overline{g}_j(x)$.

\par
   As stated above, we conclude that $f(x)^{2^k}\equiv 2\overline{g}_j(x)$ (mod $f_j(-x^{2^k})$). This implies
$$
f(x)^{2^k}=2\overline{g}_j(x) \ {\rm in} \ \mathcal{R}_j=\frac{\mathbb{Z}_4[x]}{\langle f_j(-x^{2^k})\rangle}.
$$
Here, we regard $\mathbb{F}_2[x]$ as a subset of $\mathbb{Z}_4[x]$ but $\mathbb{F}_2[x]$ is not a subring of $\mathbb{Z}_4[x]$.

\par
  Now, let $\vartheta_j(x)=\overline{g}_j(x)\in \mathcal{R}_j$. Then as a polynomial in $\mathbb{F}_2[x]$, we see that
$$\overline{g}_j(x)=\prod_{i=0}^{d_j-1}x^{2^{k-1}}(\overline{\zeta}_j^{2^i})^{2^{k-1}}
  \prod_{0\leq t\leq d_j-1, t\neq i}(x^{2^k}+(\overline{\zeta}_j^{2^t})^{2^k})$$
which satisfies
\begin{eqnarray*}
\overline{g}_j(\overline{\zeta}_j^{2^\lambda})&=&\prod_{i=0}^{d_j-1}(\overline{\zeta}_j^{2^\lambda})^{2^{k-1}}(\overline{\zeta}_j^{2^i})^{2^{k-1}}
  \prod_{0\leq t\leq d_j-1, t\neq i}((\overline{\zeta}_j^{2^\lambda})^{2^k}+(\overline{\zeta}_j^{2^t})^{2^k})\\
  &=&\overline{\zeta}_j^{2^{\lambda+k}}\prod_{0\leq t\leq d_j-1, t\neq \lambda}(\overline{\zeta}_j^{2^\lambda}-\overline{\zeta}_j^{2^t})^{2^k}.
\end{eqnarray*}
Since $\overline{\zeta}_j, \overline{\zeta}_j^2,\ldots, \overline{\zeta}_j^{2^{d_j-1}}$ are
all distinct roots in $\overline{\Gamma}_j$ of the irreducible polynomial $\overline{f}_j(x)$ and
$\overline{\zeta}_j^{2^{d_j}}=\overline{\zeta}_j$, we have $\overline{g}_j(\overline{\zeta}_j^{2^\lambda})\neq 0$
for all $\lambda=0,1,\ldots,d_j-1$. From these, we deduce that
${\rm gcd}(\overline{f}_j(x), \overline{g}_j(x))=1$. This implies ${\rm gcd}(\overline{f}_j(x)^{2^k}, \overline{g}_j(x))=1$
as polynomials in $\mathbb{F}_2[x]$. Then
by $\overline{f_j(-x^{2^k})}=\overline{f}_j(x)^{2^k}$, we see that
$f_j(-x^{2^k})$ and $\vartheta_j(x)$ are coprime in $\mathbb{Z}_4[x]$. This implies that
$a(x)\vartheta_j(x)+b(x)f_j(-x^{2^k})=1$ for some $a(x),b(x)\in \mathbb{Z}_4[x]$, i.e.
$a(x)\vartheta_j(x)\equiv 1$ (mod $f_j(-x^{2^k})$).
Hence
$\vartheta_j(x)$ is an invertible element in the ring $\mathcal{R}_j$. Therefore, we have
$\langle f_j(x)^{2^k}\rangle=\langle 2\rangle$ as ideals of $\mathcal{R}_j$.

\par
 (ii) Let $M=\langle f_j(x), 2\rangle$ be the ideal of $\mathcal{R}_j$ generated by $f_j(x)$ and $2$. Then
the residue class ring of $\mathcal{R}_j$ modulo $M$ is given by:
$$\frac{\mathcal{R}_j}{M}=\frac{\mathcal{R}_j}{\langle f_j(x), 2\rangle}\cong \frac{\mathbb{F}_2[x]/\langle \overline{f}_j(x)^{2^k}\rangle}{\langle \overline{f}_j(x)\rangle}\cong \frac{\mathbb{F}_2[x]}{\langle \overline{f}_j(x)\rangle}=\overline{\Gamma}_j,$$
where $\overline{\Gamma}_j$ is a finite field of $2^{d_j}$ elements. Hence $M$ is a maximal ideal of $\mathcal{R}_j$.
Moreover,
by (i) we have $\langle 2\rangle=\langle f_j(x)^{2^k}\rangle$. This implies
$M=\langle f_j(x)\rangle$ which is a principal ideal of $\mathcal{R}_j$ generated by $f_j(x)$.

\par
  As $f_j(x)^{2^{k+1}}=(2\vartheta_j(x))^2=4\vartheta_j(x)^2=0$ in $\mathcal{R}_j$, we see that
every element of $\mathcal{R}_j$ is nilpotent. Hence each element in $\mathcal{R}_j\setminus M$ must be an
invertible element, and so $M$ is the unique maximal ideal of  of $\mathcal{R}_j$. Therefore,
$\mathcal{R}_j$ is a finite chain ring with the unique maximal ideal $\langle f_j(x)\rangle$ by Lemma 2.1.

\par
  Let $\nu$ be the nilpotency index of $f_j(x)$ in $\mathcal{R}_j$. Then $|\mathcal{R}_j|=|\frac{\mathcal{R}_j}{\langle f_j(x)\rangle}|^\nu$ by Lemma 2.2. From this, by
$|\mathcal{R}_j|=4^{2^kd_j}$ and
$|\frac{\mathcal{R}_j}{\langle f_j(x)\rangle}|=|\frac{\mathcal{R}_j}{M}|=|\overline{\Gamma}_j|=2^{d_j}$
we deduce that
$\nu=2^{k+1}$.

\par
 (iii) Using the notation of Section 1, we know that
$\mathcal{T}_j=\{\sum_{i=0}^{d_j-1}t_ix^i\mid t_0,t_1,\ldots,t_{d_j-1}\in \{0,1\}\}\subset \mathcal{R}_j$.
As $f_j(x)$ ia a monic basic irreducible polynomial in $\mathbb{Z}_4[x]$, it follows that
$\gamma_1-\gamma_2\not\in \langle f_j(x)\rangle$ for all $\gamma_1,\gamma_2\in \mathcal{T}_j$ satisfying
$\gamma_1\neq\gamma_2$. Moreover, we have $|\frac{\mathcal{R}_j}{\langle f_j(x)\rangle}|=2^{d_j}=|\mathcal{T}_j|$.
Hence $\mathcal{T}_j$ is a system of representatives for the equivalence classes of $\mathcal{R}_j$ under
congruence modulo $f_j(x)$. Then the conclusion follows from Lemma 2.3 immediately.

\par
  (iv)--(vi) follow from properties of finite chain rings (cf. [20]). Here, we omit the proofs.
\hfill $\Box$

\vskip 3mm \par
  Next, we consider how to decompose the ring $\mathcal{A}=\frac{\mathbb{Z}_{4}[x]}{\langle x^{2^kn}+1\rangle}$
into a direct sum of finite chain rings.

\par
  Let $1\leq j\leq r$ and denote $F_j(y)=\frac{y^n-1}{f_j(y)}\in \mathbb{Z}_4[y]$. As ${\rm gcd}(\overline{F}_j(y), \overline{f}_j(y))=1$, we see that
$F_j(y)$ and $f_j(y)$ are coprime in $\mathbb{Z}_4[y]$ (cf. [26] Lemma 13.5). Hence there are polynomials $a_j(y), b_j(y)\in \mathbb{Z}_4[y]$
such that
\begin{equation}
a_j(y)F_j(y)+b_j(y)f_j(y)=1.
\end{equation}
In this paper, we define $\theta_j(x)\in\mathcal{A}$ by:

\vskip 2mm\par
 $\bullet$ $\theta_j(x)\equiv a_j(-x^{2^k})F_j(-x^{2^k})=1-b_j(-x^{2^k})f_j(-x^{2^k})$ (mod $x^{2^kn}+1$).

\vskip 2mm\noindent
   Substituting $-x^{2^k}$ for $y$ in Equations (1) of Section 1 and (2) above, we obtain
$$-(x^{2^kn}+1)=(-x^{2^k})^n-1=f_1(-x^{2^k})f_2(-x^{2^k})\ldots f_r(-x^{2^k}),$$
i.e. $x^{2^kn}+1=-f_1(-x^{2^k})f_2(-x^{2^k})\ldots f_r(-x^{2^k})$, and
$$a_j(-x^{2^k})F_j(-x^{2^k})+b_j(-x^{2^k})f_j(-x^{2^k})=1.$$
   Then from the definition of $\theta_j(x)$ and
the Chinese Remainder Theorem for commutative rings with identity,
 we deduce the following conclusion.

\vskip 3mm \noindent
   {\bf Theorem 2.6} \textit{Using the notation above, we have the following conclusions}:

\vskip 2mm \par
   (i) \textit{$\theta_1(x)+\ldots+\theta_r(x)=1$, $\theta_j(x)^2=\theta_j(x)$
and $\theta_i(x)\theta_j(x)=0$ in ${\mathcal A}$ for all $1\leq i\neq j\leq r$}.

\vskip 2mm \par
   (ii) \textit{${\mathcal A}={\mathcal A}_1\oplus\ldots\oplus {\mathcal A}_r$, where ${\mathcal A}_j=\theta_j(x){\mathcal A}$ and its multiplicative
identity is $\theta_j(x)$. Moreover, this decomposition is a direct sum of rings
in that ${\mathcal A}_i{\mathcal A}_j=\{0\}$ for all $i$ and $j$, $1\leq i\neq j\leq r$}.

\vskip 2mm \par
   (iii) \textit{For each $1\leq j\leq r$ and $a(x)\in {\mathcal R}_j=\frac{\mathbb{Z}_{4}[x]}{\langle f_j(-x^{2^k})\rangle}$, define
a map $\tau_j$ by
$$\tau_j: a(x)\mapsto \theta_j(x)a(x) \ ({\rm mod} \ x^{2^kn}+1).$$
Then $\tau_j$ is a ring isomorphism
from ${\mathcal R}_j$ onto ${\mathcal A}_j$. Hence
$|{\mathcal A}_j|=4^{2^kd_j}=2^{2^{k+1}d_j}$}.

\vskip 2mm \par
   (iv) \textit{Define $\tau: (a_1(x),\ldots,a_r(x))\mapsto \tau_1(a_1(x))+\ldots+\tau_r(a_r(x))$, i.e.
$$\tau(a_1(x),\ldots,a_r(x))=\sum_{j=1}^r\theta_j(x)a_j(x) \ ({\rm mod} \ x^{2^kn}+1)$$
for all
$a_j(x)\in {\mathcal R}_j$, $j=1,\ldots,r$. Then $\tau$ is a ring isomorphism from ${\mathcal R}_1\times\ldots$ $\times{\mathcal R}_r$ onto
${\mathcal A}$}.

\section{An explicit representation and enumeration for negacyclic codes over $\mathbb{Z}_{4}+u\mathbb{Z}_{4}$ of length $2^kn$}
In this section, we determine all distinct negacyclic codes over the ring $\mathbb{Z}_{4}+u\mathbb{Z}_{4}$ ($u^2=0$) of length $2^kn$,
i.e. all distinct ideals of the ring $\frac{(\mathbb{Z}_{4}+u\mathbb{Z}_{4})[x]}{\langle x^{2^kn}+1\rangle}$.

\par
  In this paper, for a ring $\Upsilon\in\{{\mathcal A}, {\mathcal A}_j, {\mathcal R}_j\}$ where $1\leq j\leq r$, we set $\Upsilon[u]/\langle u^2\rangle
=\Upsilon+u\Upsilon$ ($u^2=0$) in which the operations are defined by:
$$(\xi_1+u\eta_1)+(\xi_2+u\eta_2)=(\xi_1+\xi_2)+u(\eta_1+\eta_2);$$
$$(\xi_1+u\eta_1)(\xi_2+u\eta_2)=\xi_1\eta_1+u(\xi_1\eta_2+\xi_2\eta_1),$$
for any $\xi_1,\eta_1, \xi_2,\eta_2\in \Upsilon$.
  Let $\alpha\in \frac{(\mathbb{Z}_{4}+u\mathbb{Z}_{4})[x]}{\langle x^{2^kn}+1\rangle}$. Then $\alpha$ can be
expressed as
$$\alpha=\sum_{i=0}^{2^kn-1}(a_i+b_iu)x^i, \ a_i,b_i\in \mathbb{Z}_{4}, \ i=0,1,\ldots,2^kn-1$$
uniquely. Denote $\xi=\sum_{i=0}^{2^kn-1}a_ix^i$ and $\eta=\sum_{i=0}^{2^kn-1}b_ix^i$. Then we have $\xi,\eta\in {\mathcal A}
=\frac{\mathbb{Z}_{4}[x]}{\langle x^{2^kn}+1\rangle}$. Now, define $\sigma: \alpha\mapsto \xi+\eta u$.
It can be verified easily that $\sigma$ is a ring isomorphism from $\frac{(\mathbb{Z}_{4}+u\mathbb{Z}_{4})[x]}{\langle x^{2^kn}+1\rangle}$ onto
${\mathcal A}+u{\mathcal A}$.

\par
  In the rest of this paper, we will identify $\frac{(\mathbb{Z}_{4}+u\mathbb{Z}_{4})[x]}{\langle x^{2^kn}+1\rangle}$
with ${\mathcal A}+u{\mathcal A}$ under the above isomorphism $\sigma$. Moreover, we have the following.

\vskip 3mm \noindent
  {\bf Lemma 3.1} \textit{Let $1\leq j\leq r$. Using the notations of Theorem 2.6, for any
$a(x),b(x)\in {\mathcal R}_j$ we define}
\begin{eqnarray*}
\tau_j(a(x)+b(x)u) &=& \tau_j(a(x))+\tau_j(b(x))u \\
  &=&\theta_j(x)(a(x)+b(x)u) \ ({\rm mod} \ x^{2^kn}+1).
\end{eqnarray*}
\textit{Then $\tau_j$ is a ring isomorphism from ${\mathcal R}_j+u{\mathcal R}_j$ onto ${\mathcal A}_j+u{\mathcal A}_j$}.

\vskip 3mm \noindent
   \textit{Proof} By Theorem 2.6(iii), the isomorphism $\tau_j:{\mathcal R}_j\rightarrow {\mathcal A}_j$ induces
an isomorphism of polynomial rings from ${\mathcal R}_j[u]$ onto ${\mathcal A}_j[u]$ in the natural
way that
$$\sum_{i}a_i(x)u^i\mapsto \sum_{i}\tau_j(a_i(x))u^i \ (\forall a_i(x)\in {\mathcal R}_j).$$
Hence
$\tau_j$ is a ring isomorphism from ${\mathcal R}_j+u{\mathcal R}_j$ onto ${\mathcal A}_j+u{\mathcal A}_j$.
\hfill $\Box$

\vskip 3mm \noindent
  {\bf Lemma 3.2} \textit{The following statements are
equivalent}:

\vskip 2mm \par
   (i) \textit{${\mathcal C}$ is a negacyclic code over $\mathbb{Z}_{4}+u\mathbb{Z}_{4}$ of length $2^kn$}.

\vskip 2mm \par
   (ii) \textit{${\mathcal C}$ is an ideal of ${\mathcal A}+u{\mathcal A}$}.

\vskip 2mm \par
   (iii) \textit{For each integer $1\leq j\leq r$, there is a unique ideal $C_j$ of
the ring ${\mathcal R}_j+u{\mathcal R}_j$ such that}
${\mathcal C}=\bigoplus_{j=1}^r\theta_j(x)C_j \ ({\rm mod} \ x^{2^kn}+1).$
\textit{In this case, we have $|{\mathcal C}|=\prod_{j=1}^r|C_j|$}.

\vskip 3mm \noindent
 \textit{Proof} (i)$\Leftrightarrow$ (ii) It follows from the identification
of $\frac{(\mathbb{Z}_{4}+u\mathbb{Z}_{4})[x]}{\langle x^{2^kn}+1\rangle}$
with ${\mathcal A}+u{\mathcal A}$.

\par
  (ii)$\Leftrightarrow$ (iii) By Theorem 2.6(ii) we have
${\mathcal A}=\bigoplus_{j=1}^r{\mathcal A}_j$. Hence
$${\mathcal A}+u{\mathcal A}=\frac{{\mathcal A}[u]}{\langle u^2\rangle}=\bigoplus_{j=1}^r\frac{{\mathcal A}_j[u]}{\langle u^2\rangle}
=\bigoplus_{j=1}^r({\mathcal A}_j+u{\mathcal A}_j)$$
 and this decomposition is a direct sum of rings in that $({\mathcal A}_i+u{\mathcal A}_i)({\mathcal A}_j+u{\mathcal A}_j)=\{0\}$ for any $i,j$, $1\leq i\neq j\leq r$. Therefore, ${\mathcal C}$ is an ideal of ${\mathcal A}+u{\mathcal A}$
if and only if for each $1\leq j\leq r$, there is a unique ideal ${\mathcal C}_j$ of
the ring ${\mathcal A}_j+u{\mathcal A}_j$ such that ${\mathcal C}=\bigoplus_{j=1}^r{\mathcal C}_j$. From this and by Lemma 3.1, we deduce that
${\mathcal C}_j$ is an ideal of ${\mathcal A}_j+u{\mathcal A}_j$ if and only if there is a unique ideal $C_j$ of ${\mathcal R}_j+u{\mathcal R}_j$ such
that $${\mathcal C}_j=\tau_j(C_j)=\theta_j(x)C_j=\{\theta_j(x)c_j(x)\mid
c_j(x)\in C_j\} \ ({\rm mod} \ x^{2^kn}+1).$$
Then ${\mathcal C}=\bigoplus_{j=1}^r\theta_j(x)C_j$ and $|{\mathcal C}_j|=\prod_{j=1}^r|{\mathcal C}_j|=\prod_{j=1}^r|C_j|$.
\hfill $\Box$

\vskip 3mm \par
  Therefore, in order to present all nagacyclic codes over $\mathbb{Z}_{4}+u\mathbb{Z}_{4}$ of length $2^kn$,
it is sufficient to determine all ideals of
the ring ${\mathcal R}_j+u{\mathcal R}_j$ for each $j$.

\par
  Recently, by Theorem 3.8 in [13] we determined all distinct ideals of
the ring $\mathcal{K}_j+u\mathcal{K}_j$ ($u^2=0$), where
$$\mathcal{K}_j=\frac{\mathbb{F}_{p^m}[x]}{\langle f_j(x)^{p^s}\rangle}
=\{\sum_{i=0}^{p^sd_j-1}a_ix^i\mid a_0,a_1,\ldots,a_{p^sd_j-1}\in\mathbb{F}_{p^m}\}$$
in which the arithmetic is done modulo $f_j(x)^{p^s}$, and

\par
  $\diamond$ $p$ is a prime number, $m,s$ are positive integers and $\mathbb{F}_{p^m}$ is a finite field
of $p^m$ elements.

\par
  $\diamond$ $f_j(x)$ is an irreducible polynomial in $\mathbb{F}_{p^m}[x]$ with degree $d_j$.

\par
  $\diamond$ ([13] Lemma 3.1(i)) $\mathcal{K}_j$ is a finite chain ring with the unique maximal ideal
$\langle f_j(x)\rangle$ generated by $f_j(x)$, the nilpotency index of $f_j(x)$
in $\mathcal{K}_j$ is $p^s$.

\par
  $\diamond$ ([13] Lemma 3.1(ii)) Let $\mathcal{T}_j=\{\sum_{i=0}^{d_j-1}t_ix^i\mid t_0,t_1,\ldots,t_{d_j-1}\in \mathbb{F}_{p^m}\}
\subset \mathcal{K}_j$. Then each element $\xi\in \mathcal{K}_j$ has a unique $f_j(x)$-expansion:
$$\xi=\sum_{i=0}^{p^s-1}b_i(x)f_j(x)^i, \ b_0(x), b_1(x), \ldots, b_{p^s-1}(x)\in \mathcal{T}_j.$$

\par
  For clarity, we give a table below:
\begin{center}
\begin{tabular}{l|l|l}\hline
 & $\mathcal{K}_j=\frac{\mathbb{F}_{p^m}[x]}{\langle f_j(x)^{p^s}\rangle}$ in [13]
& $\mathcal{R}_j=\frac{\mathbb{Z}_4[x]}{\langle f_j(-x^{2^k})\rangle}$  \\ \hline
maximal ideal & $\langle f_j(x)\rangle$  &  $\langle f_j(x)\rangle$ \\
degree of $f_j(x)$ & $d_j$ & $d_j$ \\
nilpotency index of $f_j$ & $p^s$ & $2^{k+1}$ \\
$\mathcal{T}_j$ & $\{\sum_{i=0}^{d_j-1}t_ix^i\mid t_i\in \mathbb{F}_{p^m}\}$  & $\{\sum_{i=0}^{d_j-1}t_ix^i\mid t_i\in \mathbb{F}_{2}\}$\\
$|\mathcal{T}_j|$ & $p^{md_j}$  & $2^{d_j}$\\
elements ($b_i(x)\in \mathcal{T}_j$) & $\sum_{i=0}^{p^s-1}b_i(x)f_j(x)^i$ & $\sum_{i=0}^{2^{k+1}-1}b_i(x)f_j(x)^i$\\
\hline
\end{tabular}
\end{center}

\par
 For any positive integer $i$, let $\lceil\frac{i}{2}\rceil={\rm min}\{l\in\mathbb{Z}^{+}\mid
l\geq \frac{i}{2}\}$ and $\lfloor\frac{i}{2}\rfloor={\rm max}\{l\in\mathbb{Z}^{+}\cup\{0\}\mid
l\leq \frac{i}{2}\}$. Making the following replacements
in Theorem 3.8 of [13]:
$$\mathcal{K}_j \rightarrow \mathcal{R}_j, \ p \rightarrow 2,  \ s \rightarrow k+1,
\ k \rightarrow \lambda, \ m\rightarrow 1,$$
by Lemma 2.5 we obtain the following conclusion.

\vskip 3mm\noindent
  {\bf Theorem 3.3} \textit{Using the notations above, all distinct ideals $C_j$ of the
ring $\mathcal{R}_j+u\mathcal{R}_j$ $(u^2=0)$ are given by the following five cases:}

\vskip 2mm \par
 (I) \textit{$2^{2^kd_j}$ ideals}:

\par
  $\bullet$ \textit{$C_j=\langle f_j(x)b(x)+u\rangle$ with $|C_j|=2^{2^{k+1}d_j}$},

\noindent
  \textit{where $b(x)=\sum_{i=2^k-1}^{2^{k+1}-2}b_i(x)f_j(x)^i$ and $b_{2^k-1}(x),\ldots,b_{2^{k+1}-2}(x)\in\mathcal{T}_j$}.

\vskip 2mm \par
 (II) \textit{$\sum_{\lambda=1}^{2^{k+1}-1}2^{\left(2^{k+1}-\lambda-\lceil\frac{1}{2}(2^{k+1}-\lambda)\rceil\right)d_j}$ ideals}:

\par
  $\bullet$ \textit{$C_j=\langle uf_j(x)^{2^{k+1}-1}\rangle$ with $|C_j|=2^{d_j}$};

\par
  $\bullet$ \textit{$C_j=\langle f_j(x)^{\lambda+1}b(x)+uf_j(x)^\lambda\rangle$ with $|C_j|=2^{d_j(2^{k+1}-\lambda)}$,
where $b(x)=\sum_{i=\lceil \frac{1}{2}(2^{k+1}-\lambda)\rceil-1}^{2^{k+1}-\lambda-2}b_i(x)f_j(x)^i$,
$b_i(x)\in\mathcal{T}_j$, $\lceil \frac{1}{2}(2^{k+1}-\lambda)\rceil-1\leq i\leq 2^{k+1}-\lambda-2$
and $1\leq \lambda\leq 2^{k+1}-2$}.

\vskip 2mm \par
 (III) \textit{$2^{k+1}+1$ ideals}:

\par
  $\bullet$ \textit{$C_j=\langle f_j(x)^\lambda\rangle \ {\rm with} \ |C_j|=2^{2d_j(2^{k+1}-\lambda)}$, where $0\leq \lambda\leq 2^{k+1}.$}

\vskip 2mm \par
 (IV) \textit{$\sum_{t=1}^{2^{k+1}-1}2^{\left(t-\lceil\frac{t}{2}\rceil\right)d_j}$ ideals}:

\par
 $\bullet$ \textit{$C_j=\langle u, f_j(x)\rangle$ with $|C_j|=2^{d_j(2^{k+2}-1)}$};

\par
 $\bullet$ \textit{$C_j=\langle f_j(x)b(x)+u,f_j(x)^{t}\rangle$ with $|C_j|=2^{d_j(2^{k+2}-t)}$,
where
$b(x)=\sum_{i=\lceil \frac{t}{2}\rceil-1}^{t-2}b_i(x)f_j(x)^i$, $b_i(x)\in \mathcal{T}_j$,
$\lceil \frac{t}{2}\rceil-1\leq i\leq t-2$ and
$2\leq t\leq 2^{k+1}-1$}.

\vskip 2mm \par
 (V) \textit{$\sum_{\lambda=1}^{2^{k+1}-2}\sum_{t=1}^{2^{k+1}-\lambda-1}p^{\left(t-\lceil\frac{t}{2}\rceil\right)d_j}$ ideals}:

\par
  $\bullet$ \textit{$C_j=\langle uf_j(x)^\lambda,f_j(x)^{\lambda+1}\rangle$ with
$|C_j|=2^{d_j(2^{k+2}-2\lambda-1)}$},

\noindent
  \textit{where $1\leq \lambda\leq 2^{k+1}-1$};

\par
  $\bullet$ \textit{$C_j=\langle f_j(x)^{\lambda+1}b(x)+uf_j(x)^\lambda,f_j(x)^{\lambda+t}\rangle$
with $|C_j|=2^{d_j(2^{k+2}-2\lambda-t)}$, where
$b(x)=\sum_{i=\lceil \frac{t}{2}\rceil-1}^{t-2}b_i(x)f_j(x)^i$, $b_i(x)\in \mathcal{T}_j$,
$\lceil \frac{t}{2}\rceil-1\leq i\leq t-2$,
$2\leq t\leq 2^{k+1}-\lambda-1$ and $1\leq \lambda\leq 2^{k+1}-3$}.

\vskip 2mm \par
\textit{Moreover, let $N_{(2,d_j,2^{k+1})}$ be the number of ideals in $\mathcal{R}_j+u\mathcal{R}_j$. Then}
$$N_{(2,d_j,2^{k+1})}=\sum_{i=0}^{2^{k}}(1+4i)2^{(2^{k}-i)d_j}.$$

\vskip 3mm\noindent
  \textit{Proof} A direct proof can be given by an argument paralleling to that of Lemma 3.2, Lemma 3.7
and Theorem 3.8 in ref. [13]. Here, we omitted.
\hfill $\Box$

\vskip 3mm\par
  By the following proposition, we give a simplified expression for the
number $N_{(2,d_j,2^{k+1})}$ of ideals in $\mathcal{R}_j+u\mathcal{R}_j$.

\vskip 3mm\noindent
   {\bf Proposition 3.4} \textit{The number of ideals in $\mathcal{R}_j+u\mathcal{R}_j$ is equal to}
$$N_{(2,d_j,2^{k+1})}=\frac{(2^{d_j}+3)2^{(2^k+1)d_j}-2^{d_j}(2^{k+2}+5)+2^{k+2}+1}{(2^{d_j}-1)^2}.$$
\textit{Especially, we have $N_{(2,1,2^{k+1})}=10\cdot 2^{2^k}-2^{k+2}-9$ when $d_j=1$, and
$N_{(2,d_j,2^{1+1})}=4^{d_j}+5\cdot 2^{d_j}+9$ when $k=1$}.

\vskip 3mm\noindent
   \textit{Proof} By Theorem 3.3, it follows that
$$N_{(2,d_j,2^{k+1})}=\sum_{i=0}^{2^{k}}(2^{d_j})^{2^{k}-i}+2^{(2^k-1)d_j+2}\sum_{i=1}^{2^{k}}i(\frac{1}{2^{d_j}})^{i-1},$$
where $\sum_{i=0}^{2^{k}}(2^{d_j})^{2^{k}-i}=\sum_{l=0}^{2^{k}}(2^{d_j})^l=\frac{2^{(2^k+1)d_j}-1}{2^{d_j}-1}$. Then by
\begin{eqnarray*}
\sum_{i=1}^{2^{k}}ix^{i-1}&=&\frac{d}{dx}(\sum_{i=0}^{2^{k}}x^{i})=\frac{d}{dx}\left(\frac{x^{2^k+1}-1}{x-1}\right)\\
 &=& \frac{(2^k+1)x^{2^k}(x-1)-(x^{2^k+1}-1)}{(x-1)^2},
\end{eqnarray*}
we have
\begin{eqnarray*}
\sum_{i=1}^{2^{k}}i(\frac{1}{2^{d_j}})^{i-1}
 &=&\frac{(2^k+1)(\frac{1}{2^{d_j}})^{2^k}(\frac{1}{2^{d_j}}-1)-((\frac{1}{2^{d_j}})^{2^k+1}-1)}{(\frac{1}{2^{d_j}}-1)^2}\\
 &=&\frac{2^{-(2^k-1)d_j}}{(2^{d_j}-1)^2}\left(2^{(2^k+1)d_j}-2^{d_j}(2^k+1)+2^k\right).
\end{eqnarray*}
From these, we deduce that
\begin{eqnarray*}
N_{(2,d_j,2^{k+1})}&=&\frac{2^{(2^k+1)d_j}-1}{2^{d_j}-1}
+\frac{4}{(2^{d_j}-1)^2}\left(2^{(2^k+1)d_j}-2^{d_j}(2^k+1)+2^k\right)\\
 &=&\frac{(2^{d_j}+3)2^{(2^k+1)d_j}-2^{d_j}(2^{k+2}+5)+2^{k+2}+1}{(2^{d_j}-1)^2}.
\end{eqnarray*}
Especially, we have $N_{(2,1,2^{k+1})}=10\cdot 2^{2^k}-2^{k+2}-9$ when $d_j=1$.
\hfill $\Box$

\vskip 3mm\par
   Then from Lemma 3.2, Theorem 3.3  and Proposition 3.4, we deduce the following conclusion.

\vskip 3mm\noindent
   {\bf Theorem 3.5} \textit{Every negacyclic
code ${\mathcal C}$ over $\mathbb{Z}_{4}+u\mathbb{Z}_{4}$ $(u^2=0)$ of length $2^kn$ can be constructed by the following two steps}:

\vskip2 mm\par
  (i) \textit{For each integer $j$, $1\leq j\leq r$, choose an ideal $C_j$ of ${\mathcal R}_j+u{\mathcal R}_j$
listed in Theorem 3.3}.

\vskip2 mm\par
  (ii) \textit{Set ${\mathcal C}=\bigoplus_{j=1}^r\theta_j(x)C_j=\sum_{j=1}^r\theta_j(x)C_j$ $({\rm mod} \ x^{2^kn}+1)$.
Moreover, the number of codewords in ${\mathcal C}$ is
equal to $|{\mathcal C}|=\prod_{j=1}^r|C_j|$}.

\vskip2 mm\noindent
 \textit{Then the number of negacyclic
codes over $\mathbb{Z}_{4}+u\mathbb{Z}_{4}$ of length $2^kn$ is equal to
$$\prod_{j=1}^rN_{(2,d_j,2^{k+1})}=\prod_{j=1}^r\frac{(2^{d_j}+3)2^{(2^k+1)d_j}-2^{d_j}(2^{k+2}+5)+2^{k+2}+1}{(2^{d_j}-1)^2}.$$
Especially, the number of negacyclic
codes over $\mathbb{Z}_{4}+u\mathbb{Z}_{4}$ of length $2n$ is equal to
$\prod_{j=1}^r(4^{d_j}+5\cdot 2^{d_j}+9)$}.

\vskip 3mm\par
   Using the notations of Theorem 3.5, ${\mathcal C}=\bigoplus_{j=1}^r\theta_j(x)C_j$ is called the \textit{canonical form decomposition} of the negacyclic
code ${\mathcal C}$ over $\mathbb{Z}_{4}+u\mathbb{Z}_{4}$.

\par
  As an application of Theorem 3.5, we list the number of negacyclic
codes over $\mathbb{Z}_{4}+u\mathbb{Z}_{4}$ of length $2n$ for odd positive integers $3\leq n\leq 21$
as follows:
\begin{center}
\begin{tabular}{l|l}\hline
$n$ & The number of negacyclic
codes over $\mathbb{Z}_{4}+u\mathbb{Z}_{4}$ of length $2n$  \\ \hline
3 & $1035=(4^{1}+5\cdot 2^{1}+9)(4^{2}+5\cdot 2^{2}+9)$ \\
5 & $7935=(4^{1}+5\cdot 2^{1}+9)(4^{4}+5\cdot 2^{4}+9)$ \\
7 & $293687=(4^{1}+5\cdot 2^{1}+9)(4^{3}+5\cdot 2^{3}+9)^2$ \\
9 & $4579875=(4^{1}+5\cdot 2^{1}+9)(4^{2}+5\cdot 2^{2}+9)(4^{6}+5\cdot 2^{6}+9)$ \\
11 & $24235215=(4^{1}+5\cdot 2^{1}+9)(4^{10}+5\cdot 2^{10}+9)$ \\
13 & $386347215=(4^{1}+5\cdot 2^{1}+9)(4^{12}+5\cdot 2^{12}+9)$ \\
15 & $42500851875=(4^{1}+5\cdot 2^{1}+9)(4^{2}+5\cdot 2^{2}+9)(4^{4}+5\cdot 2^{4}+9)^4$ \\
17 & $102708354375=(4^{1}+5\cdot 2^{1}+9)(4^{8}+5\cdot 2^{8}+9)^2$ \\
19 & $1580578111695=(4^{1}+5\cdot 2^{1}+9)(4^{18}+5\cdot 2^{18}+9)$ \\
21 & $258775875646875=23\cdot 45\cdot(4^3+5\cdot2^3+9)^2(4^6+5\cdot 2^6+9)^2$ \\
\hline
\end{tabular}
\end{center}

\vskip 3mm\noindent
  {\bf Example 3.6} All distinct negacyclic codes
over $\mathbb{Z}_{4}+u\mathbb{Z}_{4}$ of length $14$ are given by:
$${\cal C}=\theta_1(x)C_1\oplus \theta_2(x)C_2\oplus \theta_3(x)C_3=\sum_{j=1}^3\theta_j(x)C_j \ ({\rm mod} \ x^{14}+1),$$
where
  \par
   $\theta_1(x)=3+x^2+3x^4+x^6+3x^8+x^{10}+3x^{12}$,

\par
   $\theta_2(x)=1+x^2+3x^4+2x^6+3x^8+2x^{10}+2x^{12}$,

\par
  $\theta_3(x)=1+2x^2+2x^4+x^6+2x^8+x^{10}+3x^{12}$;

\vskip 2mm\par
   $\bullet$ $C_1$ is one of the following $23$ distinct ideals in $\frac{(\mathbb{Z}_4+u\mathbb{Z}_4)[x]}{\langle x^{2}+1\rangle}
=\frac{(\mathbb{Z}_4+u\mathbb{Z}_4)[x]}{\langle (-x^{2})-1\rangle}$:

\par
  ({\bf 1}-I) $4$ ideals:

\par
  $\mathcal{C}=\langle (x-1)\cdot(b_1(x-1)+b_2(x-1)^2)+u\rangle$ with $|\mathcal{C}|=2^{4}$,
where $b_1,b_2\in\{0,1\}$.

\par
  ({\bf 1}-II) $5$ ideals:

\par
  $\mathcal{C}=\langle (x-1)^2\cdot b_1(x-1)+u(x-1)\rangle$ with $|\mathcal{C}|=2^{3}$, where $b_1\in\{0,1\}$;

\par
  $\mathcal{C}=\langle (x-1)^3\cdot b_0+u(x-1)^2\rangle$  with $|\mathcal{C}|=2^{2}$, where $b_0\in\{0,1\}$;

\par
  $\mathcal{C}=\langle u(x-1)^3\rangle$ with $|\mathcal{C}|=2$.

\par
  ({\bf 1}-III) $5$ ideals:

\par
  $\mathcal{C}=\langle (x-1)^\lambda \rangle$ with $|\mathcal{C}|=4^{4-\lambda}$, $0\leq \lambda\leq 4$.

\par
  ({\bf 1}-IV) $5$ ideals:

\par
  $\mathcal{C}=\langle u, x-1\rangle$ with $|\mathcal{C}|=2^{7}$;

\par
  $\mathcal{C}=\langle (x-1)\cdot b_0+u, (x-1)^2\rangle$ with $|\mathcal{C}|=2^{6}$, where $b_0\in\{0,1\}$;

\par  $\mathcal{C}=\langle (x-1)\cdot b_1(x-1)+u, (x-1)^3\rangle$ with $|\mathcal{C}|=2^{5}$,
where $b_1\in\{0,1\}$.

\par
  ({\bf 1}-V) $4$ ideals:

\par
 $\mathcal{C}=\langle u(x-1), (x-1)^2\rangle$ with $|\mathcal{C}|=2^{5}$;

\par
 $\mathcal{C}=\langle u(x-1)^2, (x-1)^3\rangle$ with $|\mathcal{C}|=2^{3}$;

\par
 $\mathcal{C}=\langle (x-1)^2\cdot b_0+u(x-1), (x-1)^3\rangle$ with $|\mathcal{C}|=2^{4}$, where $b_0\in\{0,1\}$.

\vskip 2mm\par
   $\bullet$ Denote $f_2(x)=x^3+2x^2+x+3$, $f_3(x)=x^3+3x^2+2x+3$ and
$\mathcal{T}_2=\mathcal{T}_3=\{t_0+t_1x+t_2x^2\mid t_0,t_1,t_2\in \{0,1\}\}$. Let $j=2,3$. Then $C_j$
 is one of the following $113$ ideals in $\frac{\mathbb{Z}_4[x]}{\langle f_j(-x^2)\rangle}$:

\par
  ({\bf j}-I) $64$ codes:

\par
  $\mathcal{C}=\langle f_j(x)\cdot(\beta_1f_j(x)+\beta_2f_j(x)^2)+u\rangle$ with $|\mathcal{C}|=2^{12}$,
where $\beta_1,\beta_2\in\mathcal{T}_j$.

\par
  ({\bf j}-II) $17$ codes:

\par
  $\mathcal{C}=\langle f_j(x)^2\cdot \beta_1f_j(x)+uf_j(x)\rangle$ with $|\mathcal{C}|=2^{9}$, where $\beta_1\in\mathcal{T}_j$;

\par
  $\mathcal{C}=\langle f_j(x)^3\cdot \beta_0+uf_j(x)^2\rangle$  with $|\mathcal{C}|=2^{6}$, where $\beta_0\in\mathcal{T}_j$;

\par
  $\mathcal{C}=\langle uf_j(x)^3\rangle$ with $|\mathcal{C}|=2^3$.

\par
  ({\bf j}-III) $5$ codes:

\par
  $\mathcal{C}=\langle f_j(x)^\lambda \rangle$ with $|\mathcal{C}|=4^{3(4-\lambda)}$, $0\leq \lambda\leq 4$.

\par
  ({\bf j}-IV) $17$ codes:

\par
  $\mathcal{C}=\langle u, f_j(x)\rangle$ with $|\mathcal{C}|=2^{21}$;

\par
   $\mathcal{C}=\langle f_j(x)\cdot \beta_0+u, f_j(x)^2\rangle$ with $|\mathcal{C}|=2^{18}$, where $\beta_0\in \mathcal{T}_j$

   $\mathcal{C}=\langle f_j(x)\cdot \beta_1f_j(x)+u, f_j(x)^3\rangle$ with $|\mathcal{C}|=2^{15}$,
where $\beta_1\in\mathcal{T}_j$.

\par
  ({\bf j}-V) $10$ codes:

\par
 $\mathcal{C}=\langle uf_j(x), f_j(x)^2\rangle$ with $|\mathcal{C}|=2^{15}$;

\par
 $\mathcal{C}=\langle uf_j(x)^2, f_j(x)^3\rangle$ with $|\mathcal{C}|=2^{9}$;

\par
 $\mathcal{C}=\langle f_j(x)^2\cdot \beta_0+uf_j(x), f_j(x)^3\rangle$ with $|\mathcal{C}|=2^{12}$, where $\beta_0\in\mathcal{T}_j$.

\section{Negacyclic codes over $\mathbb{Z}_{4}+u\mathbb{Z}_{4}$ of length $2^k$}
  As an application of Theorem 3.3, in this section we determine all distinct negacyclic codes
 of length $2^k$ over $\mathbb{Z}_4+u\mathbb{Z}_4$ ($u^2=0$), i.e. all distinct ideals of
the ring $\frac{(\mathbb{Z}_4+u\mathbb{Z}_4)[x]}{\langle x^{2^k}+1\rangle}$. In this case, we have
$n=1$, $f_1(x)=x-1$ with degree $d_1=1$, $r=1$, $\theta_1(x)=1$,
$\mathcal{R}_1=\frac{\mathbb{Z}_4[x]}{\langle f_1(-x^{2^k})\rangle}=\frac{\mathbb{Z}_4[x]}{\langle x^{2^k}+1\rangle}$ and $\mathcal{T}_1=\{0,1\}=\mathbb{F}_2$ as a subset of $\mathcal{R}_1$.
Then from these, by Theorem 3.3 and Proposition 3.4 we deduce the following conclusion.

\vskip 3mm \noindent
 {\bf Theorem 4.1} \textit{Let $k\geq 1$.
Then all distinct negacyclic codes
 of length $2^k$ over $\mathbb{Z}_4+u\mathbb{Z}_4$ $(u^2=0)$ are given by one of the following five cases:}

\vskip 2mm \par
 (I) \textit{$2^{2^k}$ codes}:

\par
   $\bullet$ \textit{$\mathcal{C}=\langle (x-1)b(x)+u\rangle \ {\rm with} \ |\mathcal{C}|=2^{2^{k+1}},$
where $b(x)=\sum_{i=2^k-1}^{2^{k+1}-2}b_i(x-1)^i$ with $b_i\in\{0,1\}$ for all $i=2^k-1,\ldots,2^{k+1}-2$}.

\vskip 2mm \par
 (II) \textit{$\sum_{\lambda=1}^{2^{k+1}-1}2^{2^{k+1}-\lambda-\lceil\frac{1}{2}(2^{k+1}-\lambda)\rceil}$ codes}:

\par
  $\bullet$ \textit{$\mathcal{C}=\langle u(x-1)^{2^{k+1}-1}\rangle$ with $|\mathcal{C}|=2$};

\par
  $\bullet$ \textit{$\mathcal{C}=\langle (x-1)^{\lambda+1}b(x)+u(x-1)^\lambda\rangle$  with $|\mathcal{C}|=2^{2^{k+1}-\lambda}$,
where $b(x)=\sum_{i=\lceil \frac{2^{k+1}-\lambda}{2}\rceil-1}^{2^{k+1}-\lambda-2}b_i(x-1)^i$, $b_i\in\{0,1\}$ for
all $i=\lceil \frac{2^{k+1}-\lambda}{2}\rceil-1, \ldots, 2^{k+1}-\lambda-2$,
and $1\leq \lambda\leq 2^{k+1}-2$}.

\vskip 2mm \par
 (III) \textit{$2^{k+1}+1$ codes}:

\par
  $\bullet$ \textit{$\mathcal{C}=\langle (x-1)^\lambda\rangle \ {\rm with} \ |\mathcal{C}|=2^{2(2^{k+1}-\lambda)}, \ 0\leq \lambda\leq 2^{k+1}.$}

\vskip 2mm \par
 (IV) \textit{$\sum_{t=1}^{2^{k+1}-1}2^{t-\lceil\frac{t}{2}\rceil}$ codes}:

\par
  $\bullet$ \textit{$\mathcal{C}=\langle u, (x-1)\rangle$ with $|\mathcal{C}|=2^{2^{k+2}-1}$};

\par
  $\bullet$ \textit{$\mathcal{C}=\langle (x-1)b(x)+u,(x-1)^{t}\rangle$ with $|\mathcal{C}|=2^{2^{k+2}-t}$},

\noindent
\textit{where $b(x)=\sum_{i=\lceil \frac{t}{2}\rceil-1}^{t-2}b_i(x-1)^i$, $b_i\in\{0,1\}$ for all $i=\lceil \frac{t}{2}\rceil-1, \ldots,t-2$,
and $2\leq t\leq 2^{k+1}-1$}.
\vskip 2mm \par
 (V) \textit{$\sum_{\lambda=1}^{2^{k+1}-2}\sum_{t=1}^{2^{k+1}-\lambda-1}2^{t-\lceil\frac{t}{2}\rceil}$ codes}:

\par
  $\bullet$ \textit{$\mathcal{C}=\langle u(x-1)^\lambda,(x-1)^{\lambda+1}\rangle$ with
$|\mathcal{C}|=2^{2^{k+2}-2\lambda-1}$, where $1\leq \lambda\leq 2^{k+1}-2$};

\par
  $\bullet$ \textit{$\mathcal{C}=\langle (x-1)^{\lambda+1}b(x)+u(x-1)^\lambda,(x-1)^{\lambda+t}\rangle$
with $|\mathcal{C}|=2^{2^{k+2}-2\lambda-t}$,
where
$b(x)=\sum_{i=\lceil \frac{t}{2}\rceil-1}^{t-2}b_i(x-1)^i$, $b_i\in\{0,1\}$ for all $i=\lceil \frac{t}{2}\rceil-1, \ldots,t-2$,
$2\leq t\leq 2^{k+1}-\lambda-1$ and $1\leq \lambda\leq 2^{k+1}-3$}.

\vskip 2mm \par
\textit{Moreover, the number of negacyclic codes of length $2^k$ over $\mathbb{Z}_4+u\mathbb{Z}_4$ is}
$$N_{(2,1,2^{k+1})}=10\cdot 2^{2^k}-2^{k+2}-9.$$

\par
   As in  Dougherty and Ling [19], let $h_m(x)$ be a monic basic irreducible polynomial in $\mathbb{Z}_4[x]$ of degree $m$ that
divides $x^{2^m-1}-1$ and set
$${\rm GR}(4,m)=\frac{\mathbb{Z}_4[x]}{\langle h_m(x)\rangle}=\{\sum_{i=0}^{m-1}a_ix^i\mid a_0,a_1,\ldots,a_{m-1}\in\mathbb{Z}_4\}$$
in which the arithmetic is done modulo $h_m(x)$. Then ${\rm GR}(4,m)$ is a Galois ring of characteristic $4$ and cardinality $4^m$ (cf.
[26] Theorem 14.1). Define the ring
$$R_4(z,m)=\frac{{\rm GR}(4,m)[z]}{\langle z^{2^k}-1\rangle}=\{\sum_{i=0}^{2^k-1}\alpha_i z^i\mid \alpha_0,\alpha_1,\ldots,
\alpha_{2^k-1}\in {\rm GR}(4,m)\}$$
in which the arithmetic is done modulo $z^{2^k}-1$. By Theorem 2.6 in [19], the number of
all distinct ideals in $R_4(z,m)$ is
\begin{eqnarray*}
&& 5+(2^m)^{2^{k-1}}+(5\cdot 2^m-1)(2^m)\frac{(2^m)^{2^{k-1}-1}-1}{(2^m-1)^2}-4\cdot\frac{2^{k-1}-1}{2^m-1}\\
 &=&\frac{(2^m+3)2^{(2^{k-1}+1)m}-2^m(2^{k+1}+5)+2^{k+1}+1}{(2^m-1)^2}\\
 &=& N_{(2,m,2^k)}.
\end{eqnarray*}
Especially, if $m=1$, the number all distinct ideals in $\frac{\mathbb{Z}_4[z]}{\langle z^{2^k}-1\rangle}$ is
$$N_{(2,1,2^k)}=10\cdot 2^{2^{k-1}}-2^{k+1}-9.$$
  Therefore, by Proposition 3.4 we have the following corollary.

\vskip 3mm \noindent
   {\bf Corollary 4.2} \textit{Let $1\leq j\leq r$, $f_j(x)$ be a monic basic irreducible
polynomial in $\mathbb{Z}_4[x]$ of degree $d_j$ and $\mathcal{R}_j=\frac{\mathbb{Z}_4[x]}{\langle f_j(-x^{2^k})\rangle}$ as defined in Section 1. Then the number of ideals in the ring
$\mathcal{R}_j+u\mathcal{R}_j$ $(u^2=0)$ is the same as the number
of ideals in the ring $\frac{{\rm GR}(4,d_j)[z]}{\langle z^{2^{k+1}}-1\rangle}$, where ${\rm GR}(4,d_j)=\frac{\mathbb{Z}_4[x]}{\langle f_j(x)\rangle}$. Especially, let $f_j(x)=x-1$. Then the number of ideals in the ring
$\frac{(\mathbb{Z}_4+u\mathbb{Z}_4)[x]}{\langle x^{2^k}+1\rangle}$ is the same as the number
of ideals in the ring $\frac{\mathbb{Z}_4[z]}{\langle z^{2^{k+1}}-1\rangle}$}.

\par
  \textit{Therefore, the number
of negacyclic codes over $\mathbb{Z}_4+u\mathbb{Z}_4$ with length $2^k$ is the same
as the number of cyclic codes over $\mathbb{Z}_4$  with length $2^{k+1}$}.

\vskip 3mm\par
  For examples, the number of ideals in $\mathbb{Z}_4+u\mathbb{Z}_4$, $\frac{(\mathbb{Z}_4+u\mathbb{Z}_4)[x]}{\langle x^{2}+1\rangle}$,
$\frac{(\mathbb{Z}_4+u\mathbb{Z}_4)[x]}{\langle x^{2^2}+1\rangle}$, $\frac{(\mathbb{Z}_4+u\mathbb{Z}_4)[x]}{\langle x^{2^3}+1\rangle}$,
$\frac{(\mathbb{Z}_4+u\mathbb{Z}_4)[x]}{\langle x^{2^4}+1\rangle}$, $\frac{(\mathbb{Z}_4+u\mathbb{Z}_4)[x]}{\langle x^{2^5}+1\rangle}$
and $\frac{(\mathbb{Z}_4+u\mathbb{Z}_4)[x]}{\langle x^{2^6}+1\rangle}$ is given respectively by:

\par
  $N_{(2,1,2)}=7$, $N_{(2,1,2^2)}=23$,
$N_{(2,1,2^3)}=135$, $N_{(2,1,2^4)}=2519$,

\par
  $N_{(2,1,2^5)}=655287$, $N_{(2,1,2^6)}=42949672823$,

\par
  $N_{(2,1,2^7)}=184467440737095515895$.

\vskip 3mm \noindent
  {\bf Example 4.3} All distinct $135$ negacyclic codes of length $2^2$ over $\mathbb{Z}_4+u\mathbb{Z}_4$ are given by:

\par
 (I) $2^4=16$ codes:
\par
   $\bullet$ $\mathcal{C}=\langle (x-1)\cdot(\sum_{i=3}^{6}b_i(x-1)^i)+u\rangle \ {\rm with} \ |\mathcal{C}|=2^{8},$
where $b_i\in\{0,1\}$ for all $i=3,4,5,6$.

\par
 (II) $2\cdot 2^3+2\cdot 2^2+2\cdot 2+1=29$ codes:

\par
  $\bullet$ $\mathcal{C}=\langle u(x-1)^{7}\rangle$ with $|\mathcal{C}|=2$;

\par
  $\bullet$ {$\mathcal{C}=\langle (x-1)^{\lambda+1}\cdot(\sum_{i=\lceil \frac{8-\lambda}{2}\rceil-1}^{6-\lambda}b_i(x-1)^i)+u(x-1)^\lambda\rangle$  with $|\mathcal{C}|=2^{8-\lambda}$,
where $b_i\in\{0,1\}$ for
all $i=\lceil \frac{8-\lambda}{2}\rceil-1, \ldots, 6-\lambda$,
and $1\leq \lambda\leq 6$.

\par
 (III) $9$ codes:

\par
  $\bullet$ {$\mathcal{C}=\langle (x-1)^\lambda\rangle$  with $|\mathcal{C}|=2^{2(8-\lambda)}$, where $0\leq \lambda\leq 8.$

\par
 (IV) $1+2\cdot 2+2\cdot 2^2+2\cdot 2^3=29$ codes:

\par
  $\bullet$ $\mathcal{C}=\langle u, (x-1)\rangle$ with $|\mathcal{C}|=2^{7}$;

\par
  $\bullet$ $\mathcal{C}=\langle (x-1)\cdot(\sum_{i=\lceil \frac{t}{2}\rceil-1}^{t-2}b_i(x-1)^i)+u,(x-1)^{t}\rangle$ with $|\mathcal{C}|=2^{16-t}$, where $b_i\in\{0,1\}$ for all $i=\lceil \frac{t}{2}\rceil-1, \ldots,t-2$,
and $2\leq t\leq 7$.

\par
 (V) $\sum_{\lambda=1}^{6}\sum_{t=1}^{8-\lambda-1}2^{t-\lceil\frac{t}{2}\rceil}=20+12+8+4+2+6=52$ codes:

\par
  $\bullet$ $\mathcal{C}=\langle u(x-1)^\lambda,(x-1)^{\lambda+1}\rangle$ with
$|\mathcal{C}|=2^{16-2\lambda-1}$, where $1\leq \lambda\leq 6$;

\par
  $\bullet$ $\mathcal{C}=\langle (x-1)^{\lambda+1}\cdot(\sum_{i=\lceil \frac{t}{2}\rceil-1}^{t-2}b_i(x-1)^i)+u(x-1)^\lambda,(x-1)^{\lambda+t}\rangle$
with $|\mathcal{C}|=2^{16-2\lambda-t}$, where $b_i\in\{0,1\}$ for all $i=\lceil \frac{t}{2}\rceil-1, \ldots,t-2$,
$2\leq t\leq 8-\lambda-1$ and $1\leq \lambda\leq 5$.

\vskip 3mm \noindent
  {\bf Remark 4.4} By Theorem 12 in Page 249 of ref. [4], the authors claimed that:
\textsl{The number $\mathcal{N}(2^k)$ of negacyclic codes of length $2^k$ over $R=\mathbb{Z}_4+u\mathbb{Z}_4$ is}
\begin{equation}
\mathcal{N}(2^k)=11\cdot 2^{2^k}+2^{\frac{2^k}{2}-1}(5\cdot 2^k-12)-((2^k)^2+5\cdot 2^k+4).
\end{equation}
When $k=2$, $\mathcal{N}(2^2)=176+16-40=152$  by the above formula. When $k=1$, $\mathcal{N}(2)=44-2-18=24$. In fact, this formula is incorrect.
For an example, by Example 1 in Pages
265--266 of [4], the authors claimed that ``there are
$24$ distinct codes" in $\frac{(\mathbb{Z}_4+u\mathbb{Z}_4)[x]}{\langle x^{2}+1\rangle}$ listed by the following table:
{\small \begin{center}
\begin{tabular}{ll|ll}\hline
 codes $C_i$ in ref. [4] & $|C_i|$ &  codes $C_i$ in ref. [4] & $|C_i|$\\ \hline
  $C_1=\langle 0\rangle$ & $1$ & $C_{13}=\langle (x+1)^2+u(1+(x+1))\rangle$  & $16$\\
 $C_2=\langle 1\rangle$  & $256$ & $C_{14}=\langle (x+1)^3+u\rangle$  & $16$\\
 $C_3=\langle u\rangle$  & $16$ & $C_{15}=\langle (x+1)^3+u(x+1)\rangle$  & $8$\\
 $C_4=\langle u(x+1)\rangle$   & $8$ & $C_{16}=\langle (x+1)^3+2u\rangle$  & $4$\\
 $C_5=\langle u(x+1)^2\rangle$  & $4$ &  $C_{17}=\langle (x+1),u\rangle$  & $128$\\
 $C_6=\langle u(x+1)^3\rangle$  & $2$ &  $C_{18}=\langle (x+1)^2,u\rangle$  & $64$\\
 $C_7=\langle (x+1)\rangle$  & $64$ & $C_{19}=\langle (x+1)^2,u(x+1)\rangle$  & $32$\\
 $C_8=\langle (x+1)^2\rangle$  & $16$ & $C_{20}=\langle (x+1)^2+u,u(x+1)\rangle$  & $32$\\
 $C_9=\langle (x+1)^3\rangle$  & $4$ &  $C_{21}=\langle (x+1)^3,u\rangle$  & $32$\\
 $C_{10}=\langle (x+1)+u\rangle$  & $64$ &  $C_{22}=\langle (x+1)^3,u(x+1)\rangle$  & $16$\\
 $C_{11}=\langle (x+1)^2+u\rangle$  & $16$ & $C_{23}=\langle (x+1)^3,2u\rangle$  & $8$\\
 $C_{12}=\langle (x+1)^2+u(x+1)\rangle$  & $16$ & $C_{24}=\langle (x+1)^3+2u,u(x+1)\rangle$  & $16$\\
\hline
\end{tabular}
\end{center} }

\vskip 3mm\par
  $\diamondsuit$ In this table, we note that $C_{22}=C_{24}$ and we give a proof as follows:

\par
  In the ring $\frac{(\mathbb{Z}_4+u\mathbb{Z}_4)[x]}{\langle x^2+1\rangle}$, by $x^2=-1$ and $4=0$ we have

\par
  $\bullet$ $(x-1)^2=(x+1)^2=2x$ and $2=x(x+1)^2=x(x-1)^2$.

\par
  $\bullet$ $x+1=x-1+2\in \langle x-1\rangle$
and $x-1=x+1+2\in \langle x+1\rangle$.

\par
  $\bullet$ $\langle x-1\rangle=\langle x+1\rangle$ and $(x-1)^3=(x+1)^3$.

\par
  By $2=x(x+1)^2$, we have
$$(x+1)^3+2u=(x+1)^3+x(x+1)\cdot u(x+1)\in \langle (x+1)^3, u(x+1)\rangle=C_{22}.$$
This implies $C_{24}=\langle(x+1)^3+2u,u(x+1)\rangle \subseteq C_{22}$.
  Conversely, by $$(x+1)^3=((x+1)^3+2u)-x(x+1)\cdot u(x+1)\in \langle(x+1)^3+2u, u(x+1)\rangle,$$
  it follows
that $C_{22}=\langle (x+1)^3, u(x+1)\rangle\subseteq \langle(x+1)^3+2u,u(x+1)\rangle=C_{24}$.

\par
  Therefore, $C_{22}=C_{24}$. Hence the formula (3) is wrong.

\vskip 3mm\par
   $\diamondsuit$ The uniformity of other $23$ codes in the above table and the ideals in $\frac{(\mathbb{Z}_4+u\mathbb{Z}_4)[x]}{\langle x^2+1\rangle}$ listed in ({\bf 1}-I)$\sim$({\bf 1}--V) of Example 3.6
is given by the following table. Its proof is
given in Appendix of this paper.
\begin{center}
\begin{tabular}{lll|ll}\hline
Case & codes $\mathcal{C}$ in this paper & $|\mathcal{C}|$ &  codes $C_i$ in [4] & $|C_i|$ \\ \hline
I.   &  $\langle (x-1)\cdot 0+u\rangle$  & $2^4$  & $\star$ $C_3$  & $16$\\
I.   & $\langle (x-1)\cdot((x-1)+0\cdot(x-1)^2)+u\rangle$  & $2^4$  & $\star$ $C_{11}$  & $16$\\
I.   & $\langle (x-1)\cdot((x-1)+(x-1)^2)+u\rangle$  & $2^4$  &  $\star$ $C_{13}$  & $16$\\
I.   & $\langle (x-1)\cdot(0\cdot(x-1)+(x-1)^2)+u\rangle$  &  $2^4$  & $\star$ $C_{14}$  & $16$\\
II.   &  $\langle (x-1)^2\cdot 0+u(x-1)\rangle$   & $2^3$  & $C_4$   & $8$\\
II.   &  $\langle (x-1)^3\cdot 0+u(x-1)^2\rangle$   & $2^2$  & $C_5$  & $4$\\
II.   &   $\langle u(x-1)^3\rangle$  & $2$  & $C_6$  & $2$\\
II.   & $\langle (x-1)^2\cdot(x-1)+u(x-1)\rangle$  &  $2^3$  & $C_{15}$  & $8$\\
II.   & $\langle (x-1)^3+u(x-1)^2\rangle$  &  $2^2$  & $C_{16}$  & $4$\\
III.   &  $\langle (x-1)^4\rangle$ & $1$  &  $C_1$ & $1$\\
III.   &  $\langle (x-1)^0\rangle$ & $4^4$  & $C_2$  & $256$\\
III.   & $\langle (x-1)\rangle$  & $4^3$  &$C_7$  & $64$\\
III.   & $\langle (x-1)^2\rangle$  & $4^2$  & $\star$ $C_8$  & $16$\\
III.   & $\langle (x-1)^3\rangle$  & $2^2$  &$C_9$  & $4$\\
IV.   & $\langle (x-1)+u,(x-1)^2\rangle$   & $2^6$  & $C_{10}$  & $64$\\
IV.   & $\langle u,(x-1)\rangle$  &  $2^7$  & $C_{17}$  & $128$\\
IV.   & $\langle (x-1)\cdot 0+u, (x-1)^2\rangle$  & $2^6$  &  $C_{18}$  & $64$\\
IV.   & $\langle (x-1)^2+u, (x-1)^3\rangle$  & $2^5$  &  $C_{20}$  & $32$\\
IV.   & $\langle (x-1)\cdot 0+u, (x-1)^3\rangle$  &  $2^5$  & $C_{21}$  & $32$\\
V.   & $\langle (x-1)^2+u(x-1), (x-1)^3\rangle$  & $2^4$  & $\star$ $C_{12}$  & $16$\\
V.   & $\langle (x-1)\cdot 0+u(x-1),(x-1)^3\rangle$  &  $2^4$  & $\star$ $C_{22}$  & $16$\\
V.   & $\langle u(x-1), (x-1)^2\rangle$  & $2^5$  &  $C_{19}$  & $32$\\
V.   & $\langle u(x-1)^2,(x-1)^3\rangle$  &  $2^3$  & $C_{23}$  & $8$\\
\hline
\end{tabular}
\end{center}

\vskip 3mm \noindent
  {\bf Remark 4.5} By Theorem 14 in Page 261 of [4], the authors claimed that:
\emph{The number $\mathcal{N}_S(2^k)$ of self-dual negacyclic codes of length $2^k$ over $R=\mathbb{Z}_4+u\mathbb{Z}_4$ is}
{\small
$$\mathcal{N}_S(2^k)=\frac{5\cdot 2^{2^k+2}-9\cdot 2^{e-2^k+1}+9\cdot 2^{\frac{2^k}{2}-1}
+2^{3\cdot 2^k-2e-1}(12\cdot 2^k-9e-18)+25}{9}$$ }
where $e=\lfloor (4\cdot 2^k-2)/3\rfloor$. In fact, this formula is incorrect.

\par
  For an example, the above formula claimed that there are $\mathcal{N}_S(2)=8$
distinct self-dual negacyclic codes of length $2$ over $\mathbb{Z}_4+u\mathbb{Z}_4$
(corresponding to $k=1$ and $e=2$). This result is wrong. As $C_{22}=C_{24}$, there are only
$7$ distinct self-dual negacyclic codes of length $2$ over $\mathbb{Z}_4+u\mathbb{Z}_4$:
$C_3, C_{11}, C_{13}, C_{14}, C_8, C_{12}, C_{22}$ (see the table in Page 266 of [4]).

\section{negacyclic codes of odd length over $\mathbb{Z}_4+u\mathbb{Z}_4$}
\vskip 3mm \noindent
  In previous sections, we always
suppose $k\geq 1$. In this section, let $k=0$ and we give a brief discussion on negacyclic codes of  odd length $n$ over the ring $R=\mathbb{Z}_4+u\mathbb{Z}_4$.

\par
  As in [29], we define $\phi: R^n\rightarrow \mathbb{Z}_4^{2n}$ by:
$$\phi(\xi)=(b_0,b_1,\ldots,b_{n-1},a_0+b_0,a_1+b_1,\ldots,a_{n-1}+b_{n-1})$$
for any $\xi=(a_0+ub_0,a_1+ub_1,\ldots,a_{n-1}+ub_{n-1})\in R^n$ where
$a_i,b_i\in \mathbb{Z}_4$ for all $i=0,1,\ldots,n-1$.
Then we define the Lee weight $w_L$ on $R$ by letting
$$w_L(a+ub)=w_L(b,a+b),$$
where $a,b\in \mathbb{Z}_4$ and $w_L(b,a+b)$ describes the usual Lee weight on $\mathbb{Z}_4^2$.
Furthermore, the Lee distance is defined
accordingly. Note that with this definition of the Lee weight and the Gray map we know
the following conclusion.

\vskip 3mm\noindent
  {\bf Proposition 5.1} ([29] Theorem 2.3) \textit{The map $\phi: R^n\rightarrow \mathbb{Z}_4^{2n}$ is a distance preserving linear isometry.
Thus, if $\mathcal{C}$ is a linear code over $R$ of length $n$, then $\phi(\mathcal{C})$ is a linear code over $\mathbb{Z}_4$
of length $2n$ and the two codes have the same Lee weight enumerators.}

\vskip 3mm\par
  In the following, let $n$ be odd. In the ring $\mathbb{Z}_4[x]\subset R[x]$, we have
$(-x)^n-1=-(x^n+1)$.
Hence the map $\varrho: \frac{R[x]}{\langle x^n-1\rangle}\rightarrow \frac{R[x]}{\langle x^n+1\rangle}$ defined by
$$\varrho(\alpha(x))=\alpha(-x)=\sum_{i=0}^{n-1}(-1)^i\alpha_ix^i \
(\forall \alpha(x)=\sum_{i=0}^{n-1}\alpha_ix^i \
{\rm where} \ \alpha_i\in R \ {\rm for} \ {\rm all} \ i)$$
is a ring isomorphism. Therefore, $\mathcal{C}$ is a negacyclic
code over $R$ of length $n$, i.e.
$\mathcal{C}$ is an ideal of the ring $\frac{R[x]}{\langle x^n+1\rangle}$, if and only if there is a unique cyclic code $\mathcal{D}$
 over $R$ of length $n$, i.e.
$\mathcal{D}$ is an ideal of the ring $\frac{R[x]}{\langle x^n-1\rangle}$, such that
$$\mathcal{C}=\varrho(\mathcal{D})
=\left\{\sum_{i=0}^{n-1}(-1)^i\alpha_ix^i\mid \sum_{i=0}^{n-1}\alpha_ix^i\in \mathcal{D} \ {\rm where} \
\alpha_0,\alpha_1,\ldots,\alpha_{n-1}\in R\right\}.$$
From this, one can easily verify the following conclusion.

\vskip 3mm\noindent
  {\bf Proposition 5.2} \textit{The isomorphism $\varrho$ is a distance preserving linear isometry
on $R^n$. Hence the two codes $\varrho(\mathcal{D})$ and $\mathcal{D}$ have the same Lee weight (and Hamming weight) enumerators,
for every cyclic code $\mathcal{D}$ of length $n$ over $R$}.

\vskip 3mm\par
  Therefore, it is sufficiency to determine
all cyclic codes  of length $n$ over $R=\mathbb{Z}_4+u\mathbb{Z}_4$ ($u^2=0$) in order to determine
all negacyclic codes of length $n$ over $R$ (corresponding to the case of $k=0$ in previous sections).
There were some literatures
on this kind of cyclic codes.
Please refer to [3], [21] and [29], for examples. In these papers,
the following results were given:
\begin{description}
\item{($\dag$)}
  There are $7^m$ cyclic codes of length $n$ over $R$ (Corollary 11 in [3] and Corollary 4.1 in [21]). \\
  Here $R=\mathbb{Z}_4+u\mathbb{Z}_4$ ($u^2=0$),
$x^n-1=g_1g_2\ldots g_m$ and $g_1,\ldots,g_m$ are basic irreducible
pairwise coprime polynomials in $\mathbb{Z}_4[x]\subset R[x]$.

\item{($\ddag$)}
   Let $C$ be a cyclic code of odd length $n$ over $R$. Then
$$C=\langle f_1(x)+2f_2(x)+2uf_{14}(x), uf_3(x)+2uf_4(x)\rangle,$$
where $f_2(x)\mid f_1(x)\mid x^n-1$ and $f_4(x)\mid f_3(x)\mid x^n-1$ in $R[x]$
(Theorem 4.4 in [21] and Theorem 4 in [29]).
\end{description}

\par
  Now, we provided a new way different
from the methods used in [3], [21] and [29] to study
cyclic codes of odd length $n$ over $R=\mathbb{Z}_4+u\mathbb{Z}_4$ ($u^2=0$). To do this,
we adopt the previous notations in Sections 1 and 2, for
any integer $1\leq j\leq r$ by Equations (1) and (2) we set

\vskip 2mm\par
 $\bullet$ Let $e_j(x)\in \frac{\mathbb{Z}_{4}[x]}{\langle x^n-1\rangle}\subset \frac{R[x]}{\langle x^n-1\rangle}$ be defined by
 $$e_j(x)=a_j(x)F_j(x)=1-b_j(x)f_j(x) \ ({\rm mod} \ x^n-1).$$

\par
 $\bullet$ $\mathcal{K}_j=\frac{\mathbb{Z}_{4}[x]}{\langle f_j(x)\rangle}=\{\sum_{i=0}^{d_j-1}a_ix^i\mid a_0,a_1, \ldots,a_{d_j-1}\in
\mathbb{Z}_{4}\}$ in which the arithmetic is done modulo $f_j(x)$. Then $\mathcal{K}_j$
is a Galois ring of $4^{d_j}$ elements.

\par
 $\bullet$ $\mathcal{F}_j=\{\sum_{i=0}^{d_j-1}b_ix^i\mid b_0,b_1,\ldots,b_{d_j-1}\in
\mathbb{F}_{2}\}\subset \mathcal{K}_j$.

\par
 $\bullet$ Let $\frac{\mathcal{K}_j[u]}{\langle u^2\rangle}=\mathcal{K}_j+u\mathcal{K}_j$ ($u^2=0$).

\par
 $\bullet$ Define a map $\Psi: \frac{\mathbb{Z}_{4}[x]}{\langle x^n-1\rangle}+u\frac{\mathbb{Z}_{4}[x]}{\langle x^n-1\rangle}
 \rightarrow R[x]/\langle x^n-1\rangle$ by:
for any $\xi=\sum_{i=0}^{n-1}a_{i0}x^i+u\sum_{i=0}^{n-1}a_{i1}x^i$ with $a_{i0},a_{i1}\in \mathbb{Z}_{4}$ for all $i$, we set
$$\Psi(\xi)=\sum_{i=0}^{n-1}\alpha_ix^i,
\ {\rm where} \ \alpha_i=a_{i0}+ua_{i1}\in R \
{\rm for} \ {\rm all} \ i=0,1,\ldots, n-1.$$
Then $\Psi$ is a ring isomorphism from $\frac{\mathbb{Z}_{4}[x]}{\langle x^n-1\rangle}+u\frac{\mathbb{Z}_{4}[x]}{\langle x^n-1\rangle}$
onto $R[x]/\langle x^n-1\rangle$.

\vskip 2mm\par
  In the following, we think of $\frac{\mathbb{Z}_{4}[x]}{\langle x^n-1\rangle}+u\frac{\mathbb{Z}_{4}[x]}{\langle x^n-1\rangle}$
($u^2=0$) and $R[x]/\langle x^n-1\rangle$ as the same under the isomorphism $\Psi$.
Then as a direct corollary of Theorem3 3.4 and 3.7 in [7], we obtain the following

\vskip 3mm\noindent
 {\bf Theorem 5.3} \textit{All distinct cyclic code over $R$ of odd length $n$ are given by}
$$\mathcal{C}=\bigoplus_{j=1}^re_j(x)C_j \ ({\rm mod} \ x^n-1),$$
\textit{where $C_j$ is an ideal of the ring $\mathcal{K}_j+u\mathcal{K}_j$ listed by the following table
for all $j=1,\ldots,r$}:
\begin{center}
\begin{tabular}{llll}\hline
case &  number of ideals  &  $C_j$     &   $|C_j|$ \\ \hline
I.   & $3$  & $\bullet$ $\langle u^i\rangle$ \ $(i=0,1,2)$ & $2^{2d_j(2-i)}$ \\
II.   & $2$     & $\bullet$  $\langle 2u^s\rangle$ \ $(s=0,1)$ &  $2^{d_j(2-s)}$  \\
III.   & $2^{d_j}-1$ & $\bullet$   $\langle u+2h(x)\rangle$ $(h(x)\in {\cal F}_j\setminus\{0\})$ &  $2^{2d_j}$ \\
V.   &  $1$    & $\bullet$    $\langle u,2\rangle$ &  $2^{3d_j}$ \\
\hline
\end{tabular}
\end{center}}

\noindent
\textit{Moreover, the number of codewords in $\mathcal{C}$ is equal to
$|\mathcal{C}|=\prod_{j=1}^{r}|C_j|$}.

\par
  \textit{Therefore, $\prod_{j=1}^{r}(2^{d_j}+5)$ is the number of all cyclic codes over $R$ of odd length $n$}.

\vskip 3mm\par
  For example, the number of all cyclic codes of length $7$ over $\mathbb{Z}_{4}+u\mathbb{Z}_{4}$
is $(2^1+5)\cdot (2^3+5)^2=1183$. Moreover,
we obtained $39$ formally self-dual quasi-cyclic codes of length $14$ and index $2$ over $\mathbb{Z}_4$
by the distance preserving isometry from $(\mathbb{Z}_4+u\mathbb{Z}_4)^7$ to $\mathbb{Z}_4^{14}$
 (see Section 6 of [7]).

\vskip 3mm \noindent
  {\bf Remark 5.4} In [3] and [21], the authors claimed that
the number of all cyclic codes of length $7$ over $\mathbb{Z}_{4}+u\mathbb{Z}_{4}$
is equal to $7^3=343$. In fact, the formula give in ($\dag$) is incorrect.

\par
  Using the ring isomorphism $\varrho:\frac{R[x]}{\langle x^n-1\rangle}\rightarrow \frac{R[x]}{\langle x^n+1\rangle}$,
by Theorem 5.3 one can list all distinct $\prod_{j=1}^{r}(2^{d_j}+5)$
negacyclic codes of length $n$ over $\mathbb{Z}_{4}+u\mathbb{Z}_{4}$ precisely
and easily for specific odd positive integer $n$.


\section{Conclusion}
In this paper, we give an explicit expression for every negacyclic code over $\mathbb{Z}_4+u\mathbb{Z}_4$ ($u^2=0$) of arbitrary even length $2^kn$
and provide an exact mass formula to count the number of all these codes. Moreover, we point out mistakes on the
results of representation and enumeration for negacyclic code over $\mathbb{Z}_4+u\mathbb{Z}_4$ of length $2^k$ in reference [4] and corrected them.
  A natural problem is to
represent all distinct self-dual negacyclic codes over $\mathbb{Z}_{4}+u\mathbb{Z}_{4}$ of length $2^kn$ precisely
and provide a clear formula to count the number of all these self-dual codes, for
any positive integer $k$ and odd positive integer $n$.

\begin{acknowledgements}
Part of this work was
done when Yonglin Cao was visiting Chern Institute of Mathematics, Nankai
University, Tianjin, China. Yonglin Cao would like to thank the institution
for the kind hospitality. This research is supported in part by the National
Natural Science Foundation of China (Grant Nos. 11801324, 11671235, 61571243), the Shandong Provincial Natural Science Foundation,
China (Grant No. ZR2018BA007), the Scientific Research Foundation for the PhD of Shandong University of Technology (Grant No. 417037), the Scientific Research Fund of Hubei Provincial Key Laboratory of Applied Mathematics (Hubei University)(Grant No. AM201804) and the Scientific Research Fund of Hunan
Provincial Key Laboratory of Mathematical Modeling and Analysis in
Engineering (No. 2018MMAEZD09).

\end{acknowledgements}

\section*{Appendix: Proof for the uniformity of all $23$ distinct negacyclic codes
over $\mathbb{Z}_4+u\mathbb{Z}_4$ of length $2$ and that in ref. [3].}
\noindent
By Theorem 4.1, we have the following five cases.

\vskip 2mm\par
  (I) We have the following $4$ codes:

\par
  Let $b_1=b_2=0$. Then $\mathcal{C}=\langle (x-1)\cdot 0+u\rangle=\langle u\rangle=C_3$.

\par
  Let $b_1=1$ and $b_2=0$. Then $\mathcal{C}=\langle (x-1)\cdot ((x-1)+0\cdot(x-1)^2)+u\rangle=\langle (x-1)^2+u\rangle
  =\langle (x+1)^2+u\rangle=C_{11}$.

Let $b_1=b_2=1$. Then we have
\begin{eqnarray*}
\mathcal{C} &=& \langle (x-1)\cdot((x-1)+(x-1)^2)+u\rangle=\langle (x-1)^2x+u\rangle\\
 &=&\langle (x-1)^2+u(1-(x+1))\rangle \ ({\rm by} \ x^{-1}=-x=1-(x+1))\\
 &=&\langle (x+1)^2+u(1+(x+1))+u\cdot 2(x+1) \rangle \ ({\rm by} \ -1=1+2)
\end{eqnarray*}
with $|\mathcal{C}|=2^4$. Obviously,
$$(x+1)^2+u(1+(x+1))\in C_{13}=\langle (x+1)^2+u(1+(x+1))\rangle.$$
By $u^2=0$ and $2=x(x+1)^2$, we have
$$u\cdot 2(x+1)=ux(x+1)^3=ux(x+1)\cdot((x+1)^2+u(1+(x+1)))\in C_{13}.$$
Hence $\mathcal{C}\subseteq  C_{13}$. Conversely, we have
$$(x+1)^2+u(1+(x+1))=(x-1)^2+u(1-(x+1))+u\cdot 2(x+1)$$
in which $(x-1)^2+u(1-(x+1))\in \mathcal{C}$ and
$$u\cdot 2(x+1)=u\cdot 2(x-1)=ux(x-1)^3=ux(x-1)\cdot((x-1)^2+u(1-(x+1)))\in \mathcal{C}.$$
This implies $C_{13}\subseteq \mathcal{C}$, and hence $\mathcal{C}=C_{13}$ in this case.

\par
  Let $b_1=0$ and $b_2=1$. Then $\mathcal{C}=\langle (x-1)\cdot (0\cdot(x-1)+(x-1)^2)+u\rangle=\langle (x-1)^3+u\rangle
  =\langle (x+1)^3+u\rangle=C_{14}$.

\vskip 2mm\par
  (II) We have the following $5$ codes:

\par
  Let $b_1=0$. Then $\mathcal{C}=\langle (x-1)^2\cdot 0+u(x-1)\rangle=\langle u(x-1)\rangle$.
By $u(x-1)=u(x+1)+2u=u(x+1)+x(x+1)^2u=(1+x(x+1))\cdot u(x+1)\subseteq C_4$, it follows that
$\mathcal{C}\subseteq C_4$. Symmetrically, we can prove that $C_4\subseteq \mathcal{C}$. Hence
$\mathcal{C}=C_4$.

\par
  Let $b_0=0$. Then $\mathcal{C}=\langle (x-1)^3\cdot 0+u(x-1)^2\rangle=\langle u(x-1)^2\rangle
=\langle u(x+1)^2\rangle=C_5$.

\par
  $\mathcal{C}=\langle u(x-1)^3\rangle=\langle u(x+1)^3\rangle=C_6$

\par
  Let $b_1=1$. Then $\mathcal{C}=\langle (x-1)^2\cdot (x-1)+u(x-1)\rangle=\langle (x-1)^3+u(x-1)\rangle$ with $|\mathcal{C}|=2^3$.
By $(x-1)^3=(x+1)^3$ and $(x-1)=(x+1)+2$, we have $\mathcal{C}=\langle (x+1)^3+u(x+1)+2u\rangle$.
 Obviously, $(x+1)^3+u(x+1)\in C_{15}=\langle (x+1)^3+u(x+1)\rangle$. By
$(x+1)^4=0$ and $2=x(x+1)^2$, we have $2u=x(x+1)\cdot((x+1)^3+u(x+1))\in C_{15}$. From these, we deduce that
$\mathcal{C}\subseteq C_{15}$. Symmetrically, we have $C_{15}\subseteq \mathcal{C}$.
Hence $\mathcal{C}=C_{15}$ in this case.

\par
  Let $b_0=1$. Then $\mathcal{C}=\langle (x-1)^3+u(x-1)^2\rangle=\langle (x+1)^3+u(x+1)^2\rangle$
with $|\mathcal{C}|=2^2$. By $(x-1)^2=(x+1)^2=2x$, we have
$\mathcal{C}=\langle (x+1)^3+2ux\rangle=\langle (x+1)^3+2u +2u(x+1)\rangle$.

\par
 Obviously, $(x+1)^3+2u\in C_{16}=\langle (x+1)^3+2u\rangle$. By
$(x+1)^4=0$, we have $2u(x+1)=(x+1)\cdot((x+1)^3+2u)\in C_{16}$. From these, we deduce that
$\mathcal{C}\subseteq C_{16}$. Conversely, by $(x+1)^4=2(x+1)^2=0$ it follows that
$$(x+1)^3+2u=(x+1)^3+2u +2u(x+1)+(x+1)\cdot((x+1)^3+2u +2u(x+1))\in C_{16}.$$
$\mathcal{C}=C_{16}$ in this case.

\vskip 2mm\par
  (III) We have the following $5$ codes:

\par
  Let $\lambda=0$. Then $\mathcal{C}=\langle (x-1)^4\rangle=\langle (x+1)^4\rangle=\langle 0\rangle=C_{1}$.

\par
  Let $\lambda=4$. Then $\mathcal{C}=\langle (x-1)^0\rangle=\langle (x+1)^0\rangle=\langle 1\rangle=C_{2}$.

\par
  Let $\lambda=3$. Then $\mathcal{C}=\langle (x-1)\rangle$. By $x-1=(x+1)+2=(x+1)+x(x+1)^2=(1+x(x+1))\cdot (x+1)\subseteq C_7$.
it follows that $\mathcal{C}\subseteq C_7$. Symmetrically, we have $C_{7}\subseteq \mathcal{C}$.
Hence $\mathcal{C}=C_{7}$ in this case.

\par
  Let $\lambda=2$. Then $\mathcal{C}=\langle (x-1)^2\rangle=\langle (x+1)^2\rangle=C_{8}$.

\par
  Let $\lambda=1$. Then $\mathcal{C}=\langle (x-1)^3\rangle=\langle (x+1)^3\rangle=C_{9}$.

\vskip 2mm\par
  (IV) We have the following $5$ codes:

\par
  Let $b_0=1$. Then $\mathcal{C}=\langle (x-1)+u, (x-1)^2\rangle$. By $(x+1)+u=(x-1)+u+2$ $=((x-1)+u)+x(x-1)^2$, we have $C_{10}=\langle (x+1)+u\rangle\subseteq \mathcal{C}$.
Conversely, by
$$(x-1)^2=2x=(x+1)^2, \ (x-1)+u=(x+1)+u+2=((x+1)+u)+x(x+1)^2$$
 and $(x+1)^2=((x+1)+u)((x+1)-u)\in \langle(x+1)+u\rangle$,
we have that $(x-1)+u, (x-1)^2\in \langle(x+1)+u\rangle$. This implies $\mathcal{C}\subseteq C_{10}$, and so
$\mathcal{C}=C_{10}$.

\par
 $\mathcal{C}=\langle u, x-1\rangle$. Then by $x-1=(x+1)+2=(1+x(x+1))\cdot (x+1)\in \langle (x+1),u\rangle=C_{17}$,
it follows that $\mathcal{C}\subseteq C_{17}$. Symmetrically, we have $C_{17}\subseteq \mathcal{C}$.
Hence $\mathcal{C}=C_{17}$ in this case.

\par
  Let $b_0=0$.  Then
$$\mathcal{C}=\langle (x-1)\cdot 0+u, (x-1)^2\rangle=\langle u, (x-1)^2\rangle=\langle (x+1)^2, u\rangle=C_{18}.$$

\par
  Let $b_0=1$.  Then $\mathcal{C}=\langle (x-1)+u, (x-1)^2\rangle=\langle (x-1)+u, (x+1)^2\rangle$.
By $(x-1)+u=(x+1)+u+2=((x+1)+u)+x\cdot(x+1)^2\in \langle (x+1)^2, (x+1)+u\rangle$ $=C_{19}$, it folows
that $\mathcal{C}\subseteq C_{19}$. Symmetrically, we have $C_{19}\subseteq \mathcal{C}$.
Hence $\mathcal{C}=C_{19}$ in this case.

\par
  Let $b_1=1$.  Then $\mathcal{C}=\langle (x-1)\cdot (x-1)+u, u(x-1)\rangle=\langle (x-1)^2+u, u(x-1)\rangle
=\langle (x+1)^2+u, u(x-1)\rangle$. By $u(x-1)=u(x+1)+2u=u(x+1)+2x(x+1)^2 u$ $=(1+x(x+1))\cdot u(x+1)\in C_{10}$, it follows
that $\mathcal{C}\subseteq \langle (x+1)^2+u, u(x+1)\rangle=C_{20}$. Symmetrically, we have $C_{20}\subseteq \mathcal{C}$.
Hence $\mathcal{C}=C_{20}$ in this case.

\par
 (V) We have the following $4$ codes:

\vskip 2mm\par
  Let $b_0=1$. Then $\mathcal{C}=\langle (x-1)^2+u(x-1),(x-1)^3\rangle$. By
$(x-1)^3=(x-1)\cdot((x-1)^2+u(x-1))-u(x-1)^2$ and $u(x-1)^2=u\cdot((x-1)^2+u(x-1))$, it follows that
$$\mathcal{C}=\langle (x-1)^2+u(x-1)\rangle=\langle (x+1)^2+u(x+1)+2u\rangle.$$
Obviously, $(x+1)^2+u(x+1)\in C_{12}=\langle (x+1)^2+u(x+1)\rangle$. Then by
$2u=x(x+1)^2u=xu\cdot ((x+1)^2+u(x+1))\in C_{12}$, we have $\mathcal{C}\subseteq C_{12}$.
Symmetrically, we have $C_{12}\subseteq \mathcal{C}=\langle (x-1)^2+u(x-1)\rangle$.
Hence $\mathcal{C}=C_{12}$ in this case.

\par
  Let $b_0=0$. Then
  $$\mathcal{C}=\langle (x-1)^2\cdot 0+u(x-1),(x-1)^3\rangle=\langle u(x-1),(x-1)^3\rangle=\langle u(x-1),(x+1)^3\rangle.$$
   By $u(x-1)=(1+x(x+1))\cdot u(x+1)\in \langle(x+1)^3, u(x+1)\rangle=C_{22}$,
it follows that $\mathcal{C}\subseteq C_{22}$.
Symmetrically, we have $C_{22}\subseteq \mathcal{C}$.
Hence $\mathcal{C}=C_{22}$.

\par
   $\mathcal{C}=\langle u(x-1), (x-1)^2\rangle=\langle u(x-1), (x+1)^2\rangle$. Then by
$$u(x-1)=(1+x(x+1))\cdot u(x+1)\in \langle(x+1)^2, u(x+1)\rangle=C_{22},$$
it follows that $\mathcal{C}\subseteq C_{19}$.
Symmetrically, we have $C_{19}\subseteq \mathcal{C}$.
Hence $\mathcal{C}=C_{19}$.

\par
   $\mathcal{C}=\langle u(x-1)^2, (x-1)^3\rangle=\langle u(x+1)^2, (x+1)^3\rangle$. By
$$u(x+1)^2=u\cdot 2x=x\cdot 2u\in C_{23}=\langle (x+1)^3,2u\rangle,$$
it follows that $\mathcal{C}\subseteq C_{23}$.
Symmetrically, we have $C_{23}\subseteq \mathcal{C}$.
Hence $\mathcal{C}=C_{23}$.


\begin{thebibliography}{s99}
%
%

\bibitem{s1} Abualrub, T., Oehmke, R.: On the generators of
$\mathbb{Z}_4$ cyclic codes of lenth $2^e$, IEEE Trans. Inform.
Theory {\bf 49} (2003), 2126--2133.

\bibitem{s2} Abualrub, T., Siap, I.: Cyclic codes over the ring $\mathbb{Z}_2+u\mathbb{Z}_2$ and $\mathbb{Z}_2+u\mathbb{Z}_2+u^2\mathbb{Z}_2$,
Des. Codes Cryptogr. {\bf 42}, 273--287 (2007).

\bibitem{s3} Bandi, R.K., Bhaintwal, M.: Cyclic codes over $\mathbb{Z}_4+u\mathbb{Z}_4$, \\
https://www.researchgate.net/publication/289506486

\bibitem{s4} Bandi, R.K., Bhaintwal, M., Aydin, N: A mass formula for negacyclic codes
of length $2^k$ and some good negacyclic codes over $\mathbb{Z}_4+u\mathbb{Z}_4$,
Cryptogr. Commun. {\bf 9} (2017), 241--272.

\bibitem{s5} Blackford, T.: Negacyclic codes over $\mathbb{Z}_4$ of
even length, IEEE Trans. Inform. Theory {\bf 49} (2003), 1417--1424.

\bibitem{s6} Cao, Y.:  On constacyclic codes over finite chain rings, Finite
Fields Appl. {\bf 24} (2013), 124--135.

\bibitem{s7} Cao, Y., Li, Q.: Cyclic codes of odd length over $\mathbb{Z}_4[u]/\langle u^k\rangle$,
Cryptogr. Commun. {\bf 9} (2017), 599--624.

\bibitem{s8} Cao, Y., Cao, Y., Fu, F.-W.: Cyclic codes over $\mathbb{F}_{2^m}[u]/\langle u^k \rangle$ of oddly even length,
Appl. Algebra in Engrg. Commun. Comput. {\bf 27} (2016), 259--277.

\bibitem{s9} Cao, Y., Cao, Y., Li, Q.: Concatenated structure of cyclic codes over $\mathbb{Z}_4$ of length $4n$,
Appl. Algebra in Engrg. Commun. Comput. {\bf 27} (2016), 279--302.

\bibitem{s10} Cao, Y., Cao, Y., Li, Q.: The concatenated structure of cyclic codes over $\mathbb{Z}_{p^2}$,
J. Appl. Math. Comput. {\bf 52} (2016), 363--385.

\bibitem{s11} Cao, Y., Cao, Y.: Complete classification for simple-root cyclic codes over $\mathbb{Z}_{p^s}[v]/\langle v^2-pv\rangle$,
https://www.researchgate.net/publication/320620031

\bibitem{s12} Cao Y., Cao Y.: Negacyclic codes over the local ring $\mathbb{Z}_4[v]/\langle v^2+2v\rangle$
of oddly even length and their Gray images, Finite
Fields Appl. {\bf 52} (2018), 67--93.

\bibitem{s13} Cao, Y., Cao, Y., Dinh, H. Q., Fu, F-W., Gao, J.,  Sriboonchitta, S.:  Constacyclic codes of length $np^s$ over $\mathbb{F}_{p^m} + u\mathbb{F}_{p^m} $, Adv. Math. Commun. {\bf 12} (2018), 231--262.

\bibitem{s14} Cao, Y., Cao, Y.,  Bandi, R. K., Fu, F-W:
An explicit representation and enumeration for negacyclic codes of length $2^kn$ over $\mathbb{Z}_4+u\mathbb{Z}_4$,  http://arxiv.org/submit/2486441/pdf

\bibitem{s15} Dinh, H. Q., L\'{o}pez-Permouth, S. R.:
Cyclic and negacyclic codes over finite chain rings, IEEE Trans.
Inform. Theory {\bf 50} (2004), 1728--1744.

\bibitem{s16} Dinh, H. Q.: Constacyclic codes of length $p^s$ over
$\mathbb{F}_{p^m}+u \mathbb{F}_{p^m}$, J. Algebra, {\bf 324} (2010),
940--950.

\bibitem{s17} Dinh, H. Q., Dhompongsa, S., Sriboonchitta, S.: Repeated-root constacyclic codes of prime power length over  $\frac{\mathbb{F}_{p^m}[u]}{\langle u^a\rangle}$
and their duals, Discrete Math. {\bf 339} (2016), 1706--1715.

\bibitem{s18} Dougherty S. T., Kim J-L, Kulosman H., Liu H.: Self-dual
codes over commutative Frobenius rings. Finite Fields Appl. {\bf 16} (2010), 14--26.

\bibitem{s19} Dougherty, S. T., Ling, S.: Cyclic codes over $\mathbb{Z}_4$ of even length,
Des. Codes Cryptogr. {\bf 39} (2006), 127--153.

\bibitem{s20} Norton, G.,  S\u{a}l\u{a}gean-Mandache, A: On the structure of linear and cyclic
codes over finite chain rings, Appl. Algebra in Engrg. Comm. Comput.
{\bf 10} (2000), 489--506.

\bibitem{s21} Pattanayak, P., Singh, A. K.: A class of cyclic codes cver the Ring
$\mathbb{Z}_4[u]/\langle u^2\rangle$ and its Gray image,
http://arxiv.org/abs/1507.04938v1

\bibitem{s22} S\u{a}l\u{a}gean, A: Repeated-root cyclic and
negacyclic codes over finite chain rings. Discrete Appl. Math. {\bf
154} (2006), 413--419.

\bibitem{s23} Shi, M., Xu, L., Yang, G.: A note on one weight and two weight projective $\mathbb{Z}_4$-codes, IEEE Trans.
Inform. Theory {\bf 63}(1) (2017), 177--182.

\bibitem{s24} Shi, M., Sol\'{e}, P., Wu, B.: Cyclic codes and the weight enumerators over
$\mathbb{F}_2 +v\mathbb{F}_2 +v^2\mathbb{F}_2$, Applied and Computational Mathematics {\bf 12}(2) (2013),
247--255.

\bibitem{s25} Shi, M., Qian, L., Sok, L.,  Aydin, N.,  Sol\'{e}, P.:
On constacyclic codes over $\mathbb{Z}_4[u]/\langle u^2-1\rangle$ and
their Gray images, Finite Fields Appl. {\bf 45}:3 (2017), 86--95.

\bibitem{s26} Wan Z.-X.:  Lectures on finite fields and Galois
rings. World Scientific Pub Co Inc. 2003.

\bibitem{s27} Wood, J. A.: Duality for modules over finite rings and applications to coding theory,
American Journal of Mathematics, {\bf 121}(3) (1999), 555--575.

\bibitem{s28} Yildiz, B., Karadeniz, S.: Linear codes over $\mathbb{Z}_4+u\mathbb{Z}_4$: MacWilliams identities, projections, and formally self-dual codes.
Finite Fields Appl. {\bf 27} (2014), 24--40.

\bibitem{s29} Yildiz, B. and Aydin, N.:  Cyclic codes over $\mathbb{Z}_4+u\mathbb{Z}_4$ and
$\mathbb{Z}_4$ images, International Journal
of Information and Coding Theory, {\bf 2}(4) (2014), 226--237.
\end{thebibliography}


 \end{document}